\documentclass[letterpaper]{emulateapj}
\usepackage{amsmath}
\usepackage{natbib}
\usepackage{lscape}

\shorttitle{}
\shortauthors{Graves et al.}

\begin{document}

\title{Dissecting the Red Sequence---I. Star Formation Histories of
  Quiescent Galaxies: The Color-Magnitude vs. the Color-$\sigma$
  Relation}

\author{Genevieve J. Graves\altaffilmark{1},
S. M. Faber\altaffilmark{1}, \& Ricardo P. Schiavon\altaffilmark{2}}

\altaffiltext{1}{UCO/Lick Observatory, Department of Astronomy and
  Astrophysics, University of California, Santa Cruz, CA 95064}
\altaffiltext{2}{Gemini Observatory, 670 N. A'ohoku Place, Hilo, HI 96720}


\begin{abstract}

We use a sample of $\sim$16,000 non-emission line galaxies from the
SDSS to investigate the physical parameters underlying the well-known
color-magnitude and color-$\sigma$ relations.  Galaxies are sorted in
terms of velocity dispersions ($\sigma$), luminosity ($L$), and color,
and their spectra are stacked to obtain very high $S/N$ mean spectra
for stellar population analysis.  This allows us to map mean
luminosity-weighted ages, [Fe/H], [Mg/H], and [Mg/Fe] in
$\sigma$-$L$-color space.  Our first result is that there are many
different red sequences, with age, [Fe/H], [Mg/H], and [Mg/Fe] showing
different amounts of slope and scatter when plotted versus $\sigma$,
$L$, or color.  These behaviors are explained if the star formation
histories of the galaxies populate a two-dimensional parameter space.
One parameter is the previously well-known increase in age, [Fe/H],
[Mg/H], and [Mg/Fe] with $\sigma$.  In addition to this, we find
systematic variations at fixed $\sigma$, such that more luminous
galaxies are younger, more Fe-rich, but have lower [Mg/Fe] than their
fainter counterparts.  The main $\sigma$ trends support a paradigm in
which more massive galaxies form their stars more rapidly and at
earlier times than less massive galaxies.  The trends at fixed
$\sigma$ are consistent with scatter in the duration of star formation
for galaxies at a given $\sigma$.  The co-variation of stellar
population properties and $L$ residuals at fixed $\sigma$ that we
present here has a number of implications: it explains the differing
behavior of stellar population indicators when investigated versus
$\sigma$ as compared to $L$, and it reveals that $L$ is not as
efficient as $\sigma$ for indicating galaxy ``size'' in stellar
population studies.

\end{abstract}

\section{Introduction}\label{introduction}

Early type galaxies are observed to obey many scaling relations
between their structural and spectral properties.  Early work
identified a number of one-dimensional relations, including a
color-magnitude relation \citep{faber73,sandage78,bower92}, color-line
strength relations \citep{faber73}, the Faber-Jackson relation between
galaxy luminosity ($L$) and velocity dispersion ($\sigma$)
\citep{faber76}, variations in galaxy mass-to-light ratio ($M/L$)
versus $L$ \citep{tinsley81,faber87}, and various correlations of
$\sigma$ and $L$ with galaxy effective radius ($R_e$) and effective
surface brightness ($\mu_e$) \citep{kormendy85}, or with galaxy core
radius ($r_c$) and central surface brightness ($I_0$) \citep{lauer85}.
It is clear however that early type galaxies actually comprise at
least a two-parameter population in terms of their structure, as
represented by the Fundamental Plane (FP) relation
\citep{djorgovski87, dressler87}.  Projections of the FP appear narrow
in some orientations, leading to the seemingly one-dimensional
relations listed above.

In comparison to structure, our understanding of the stellar
population scaling laws is less well developed and has primarily
focused on one-dimensional relations, despite the evidence that early
type galaxies form a two-parameter population, at least structurally.
The stellar populations of early type galaxies are known to vary with
galaxy ``size'', where size is typically parameterized as either
$\sigma$ or $L$, but sometimes also as stellar mass ($M_*$).  Such
relations include the color-magnitude and color-$\sigma$ relations
(where it is assumed that the galaxy color is indicative of its
stellar population) as well as absorption line strength-$\sigma$
relations (e.g.,
\citealt{burstein84,bender93,trager98,colless99,kuntschner01,bernardi03d}),
which give potentially more detailed information about the stellar
populations than do broad-band colors.  In general, colors are
observed to redden and metal line strengths to increase with
increasing $\sigma$ or $L$.

The question of whether these variations are due primarily to
variations in metallicity, or whether age variations may also play a
role was first posed by \citet{faber73} and is still debated in the
literature.  The age-metallicity degeneracy is such that differences
in population age and differences in population metallicity affect
colors and metal absorption lines in nearly indistinguishable ways.
This can be loosely quantified by the ``3/2'' rule, that a fractional
change in metallicity is roughly equivalent to a 1.5 times larger
fractional change in age in terms of its effect on colors and metal
lines \citep{worthey94_models}.  Many authors (e.g.,
\citealt{bower92,kodama97}) have since claimed that the
color-magnitude relation is driven primarily by metallicity variations
rather than age variations, although \citet{faber95} argued that age
effects were also important.  Recently, \citet{gallazzi06} showed that
galaxy metallicities increase by 100\% along the red sequence, while
ages increase by only 50\%.  Applying the 3/2 rule, metallicity
variations should therefore account for $\sim75$\% of the observed
color variation along the color-magnitude relation, while age
variation provides the remaining $\sim25$\%.

Other recent work studying stellar population trends with $\sigma$
(rather than $L$) have demonstrated age variations with $\sigma$ that
should make larger contributions to the color variation.  Applying the
3/2 rule to these studies, age appears to contribute from $\sim 33$\%
\citep{thomas05, graves07} to $\sim 40$\%
\citep{trager00b,nelan05,smith07} of the observed color variation
along the color-$\sigma$ relation.

In this paper, we tackle these subtle differences head-on, comparing
the color-magnitude and color-$\sigma$ relations directly.  We explore
in detail the related stellar population variations, focusing on age,
metallicity, and abundance ratio variations with $\sigma$ and
separately with $L$.  We then look at how stellar populations
correlate with residuals from the mean $\sigma$-$L$ relation.  This
analysis explains how these systematic residual relations conspire
to produce the similar yet subtly and confusingly different observed
trends with $\sigma$ and $L$.

This is the first in a series of papers presenting a detailed
exploration of stellar population properties of quiescent (non-star
forming) galaxies, as well as the dependence of these properties on
the structural parameters of the galaxies.  The ultimate goal is to
match up information about the star formation histories of quiescent
galaxies (derived from stellar population analysis) to the mass
assembly and structural evolution of these galaxies (as inferred from
galaxy structure and morphology).  We use a sample of $\sim$16,000
quiescent galaxies from the Sloan Digital Sky Survey (SDSS), stacking
dozens or hundreds of similar galaxies to get very high
signal-to-noise ($S/N$) spectra for detailed and accurate stellar
population work.  For the purpose of this analysis, we differentiate
between ``global variables'', which we use to identify similar
galaxies for stacking, and ``stellar population variables'', which are
the mean ages, metal abundances, and abundance ratios derived from the
stacked spectra.

This first paper focuses on the color-magnitude and color-$\sigma$
relations of quiescent galaxies; the global variables considered are
therefore $\sigma$, $L$, and color\footnote{It is worth keeping in
mind that, although $L$ and color are treated as ``global'' variables
in this work, they (unlike $\sigma$) depend significantly on the
stellar population properties of the galaxy and are thus not truly
independent {\it structural} parameters.  We call them ``global''
(along with $\sigma$) for linguistic simplicity.}.  The goal is to
understand where these relations come from, what causes scatter around
the relations, and how and why the color-magnitude and color-$\sigma$
relations differ from one another.  The color-magnitude relation is a
natural place to begin this series of papers, as it is the most
accepted and most widely used of the early type galaxy scaling laws,
to the extent that it is even sometimes used to identify galaxy
clusters in photometric surveys where no redshifts are available
(e.g., \citealt{gladders00}).  The stellar population variables
presented here include mean stellar age, total Fe abundance ([Fe/H]),
total Mg abundance ([Mg/H]), and the relative abundance ratio of Mg
and Fe ([Mg/Fe]).

Much work has already been done on these topics and not all of the
results presented in this paper are new.  However, the origin of the
color-magnitude relation is fundamental to understanding the formation
and evolution of early type galaxies, yet as discussed above, there is
considerable confusion in the literature as to how the various scaling
laws fit together.  By directly investigating the various relations
between $\sigma$, $L$, color, and stellar population properties
simultaneously, we show that the various stellar population variables
correlate quite strongly with some of the global variables but only
very weakly with other global variables.  In particular, $\sigma$ and
$L$ are not interchangeable.  For instance, age correlates strongly
with $\sigma$ but rather weakly with $L$, while [Fe/H] shows the
opposite behavior.

Briefly looking forward to the subsequent papers in this series, Paper
II maps stellar population variations throughout the Fundamental Plane
(FP).  In Paper II we show that the stellar populations of galaxies on
the FP vary only with $\sigma$; the second dimension of the face-on FP
appears to be uncorrelated with the galaxy star formation histories.
However, variations are observed through the {\it thickness} of the
FP.  Paper III demonstrates that the 2-D family of quiescent galaxy
stellar populations can be mapped onto a 2-D cross-section through the
FP and presents a toy model of galaxy star formation histories.
Finally, Paper IV compares various stellar ($M_{\star}/L$) and
dynamical ($M_{dyn}/L$) mass-to-light ratios for the galaxy sample.
Paper IV demonstrates that the variations in $M_{\star}/L$ due to
stellar population effects are inadequate to explain the observed
variation in $M_{dyn}/L$, both along the FP and perpendicular to the
plane.

In \S\ref{sample}, we describe the sample of galaxies used in this
analysis.  Section \ref{cmr} reviews known properties of the color
magnitude relation for early type galaxies and the distribution of
these galaxies in the three dimensional $\sigma$-$L$-color space as a
motivation for the subsequent analysis.  In
\S\ref{deconstructing_cmr}, we discuss the process by which we
deconstruct the color-magnitude relation and measure stellar
population properties {\it across} the red sequence.  Section
\ref{stellar_pops} presents the results of the analysis and maps
stellar population ages, [Fe/H], [Mg/H] and $\alpha$-enhancement in
the three-dimensional $\sigma$-$L$-color space.  This is followed by a
brief discussion of the results in \S\ref{discussion}, and a summary
of our conclusions in \S\ref{conclusions}.

\section{The Quiescent Galaxy Sample}\label{sample}

The goal of this analysis is to achieve a fundamental understanding of
the correlations between $\sigma$, $L$, and color for typical early
type galaxies and their relation to stellar population properties.
However, from the very outset we face challenges in defining a
reasonable sample with which to explore these relations.  Although
there is a broad correspondance between galaxies with early type
morphologies, galaxies that lie on the red sequence, and galaxies that
are not actively forming stars, not all early type galaxies are on the
red sequence, not all red sequence galaxies have early type
morphology, and some early type galaxies are still actively forming
stars.

\citet{schawinski07} identify a sample of {\it morphologically
  selected} SDSS early type galaxies and classify their emission line
  properties using the emission line ratio diagrams pioneered by
  \citet[hereafter BPT]{baldwin81}.  Their early type sample consists
  of 81.5\% quiescent galaxies (i.e., no $>3\sigma$ emission
  detections in H$\alpha$, H$\beta$, [O\textsc{iii}]$\lambda$3727, or
  [N\textsc{ii}]$\lambda$6583), 1.5\% Seyferts, 5.7\% low ionization
  nuclear emission-line regions (LINERs), and 11.2\% star-forming or
  transition objects (which host both star formation and AGN
  activity).  The majority of their star-forming galaxies, transition
  objects, and Seyferts are not on the red sequence; clearly it would
  not make sense to include such objects in an analysis of the
  color-magnitude and color-$\sigma$ relations for passive galaxies.
  \citet{schawinski07} require $>3\sigma$ detections in all four
  emission lines listed above and therefore likely identify as
  ``quiescent'' numerous galaxies with low-level emission that may be
  detected in the strongest emission line but is below the $3\sigma$
  threshold in weaker lines such as H$\beta$ or [N\textsc{ii}].

In a parallel study, \citet{yan06} classify a volume-limited
color-selected sample of SDSS red sequence galaxies based on their
emission line ratios, also using BPT diagrams.  They identify as
quiescent only galaxies that have {\it no} emission lines detected at
the $3\sigma$ level.  In practice, the strong H$\alpha$ and
[O\textsc{ii}]$\lambda$3727 emission lines drive this definition.
They find that only 47.8\% of red sequence galaxies are quiescent,
with a further 28.8\% exhibiting LINER-like emission and 23.4\% likely
hosting Seyfert activity, star formation, or a combination of the two.
Their red sequence galaxies almost certainly include dust-reddened
star-forming galaxies, who dust-corrected colors would lie off of the
red sequence.  

In light of these studies, we have chosen to study the correlations
between $\sigma$, $L$, color, and stellar populations for a sample of
{\it quiescent} galaxies.  Excluding galaxies with active star
formation removes potential contamination from dust-reddened
star-forming objects whose colors do not reflect their underlying
stellar populations.  The quiescent criterion also excludes Seyferts
and LINERs.  Many Seyferts have early type morphologies but typically
have substantially younger stellar populations \citep{kauffmann03}, to
the extent that they mostly do not fall on the red sequence
\citep{schawinski07} and are not relevant to the typical early type
galaxy color-magnitude and color-$\sigma$ relations.  Furthermore, a
strong point source may contaminate global galaxy color
determinations.

A substantially stronger case could be made for including galaxies
with LINER emission because they are common on the red sequence and
typically reside in early type hosts with red colors.  In
\citet{graves07}, we analysed the stellar populations of
non-star-forming red sequence galaxies, comparing quiescent galaxies
with those hosting LINERs.  We found that the LINER hosts had
systematically younger ages, but the same metallicities and abundance
patterns as quiescent red sequence galaxies at the same $\sigma$.
They also have similar or slightly {\it redder} colors than their
quiescent counterpart, suggesting that the color effects of their
younger populations are offset by larger mean dust content.  Excluding
them from the sample in this analysis will remove many of the most
recent arrivals on the red sequence.

However, we have decided to limit the analysis in this work to
quiescent galaxies only, saving the LINERs for a future separate
analysis.  It can be difficult to distinguish between LINERs and
transition objects \citep{schawinski07}.  There is no real consensus
in the literature as to the true nature of LINER sources, which may
well represent a heterogeneous populations.  Furthermore, there is
some indication that the LINER sample of \citet{graves07} in fact
contains two separate sub-populations.  A preliminary analysis
suggests that these sub-populations have different stellar population
properties from one another and from the quiescent galaxy population,
as well as different luminosity and mass distributions at fixed
$\sigma$.  As such, the LINER population warrant a separate and
detailed analysis.

The sample discussed in this paper is therefore a sample of quiescent
galaxies, chosen spectroscopically to be free of emission lines of any
kind.  It is not a representative sample of ``early type galaxies'' or
``red sequence galaxies'', but instead provides a window onto the
stellar population properties and star formation histories of galaxies
that are no longer forming stars.  The reader should bear in mind that
the exclusion of LINERs from the sample may remove many galaxies whose
star formation ended very recently and that the mean ages of the
galaxies in this analysis are therefore biased toward older values.
If the stellar population trends of the LINER host galaxies follow the
same behavior as quiescent galaxies, the {\it relative} properties
presented here should hold for the non-star-forming galaxy population
as a whole even if the zeropoint of the age scale is uncertain.
However, the properties of the LINER population must be explored in
future work to verify this possibility.

\subsection{SDSS Data}\label{sdss_data}

The galaxies used in this analysis are drawn from the Sloan Digital
Sky Survey (SDSS, \citealt{york00}), which was conducted with a
dedicated 2.5m telescope at Apache Point Observatory \citep{gunn06}.
Images are taken in drift-scanning mode with a mosaic CCD camera
\citep{gunn98}, processed \citep{lupton01, stoughton02, pier03}, and
calibrated \citep{hogg01, smith02, ivezic04, tucker06}.  All galaxies
used in this work are part of the spectroscopic Main Galaxy Sample
\citep{strauss02}.  

Redshifts ($z$), de Vaucouleurs model photometry ($ugriz$ apparent
magnitudes, \citealt{fukugita96}), Galactic extinctions, and Petrosian
radii enclosing 50\% and 90\% of the total galaxy light ($R_{50}$ and
$R_{90}$, respectively) are taken from the NYU Value-Added Galaxy
Catalog \citep{blanton05-vagc} version of SDSS Data Release Four
\citep{adelman-mccarthy06}.  These are supplemented with measurements
from the SDSS catalog archive
server\footnote{http://cas.sdss.org/dr6/en/} of velocity dispersions
($\sigma_{fib}$, as measured in the SDSS fiber aperture), $r$-band de
Vaucouleurs radii ($R_{deV}$) and de Vaucouleurs profile axis ratios
($(a/b)_{deV}$), as well as the likelihoods of de Vaucouleurs and
exponential fits to the galaxy light profiles.  The catalog archive
server also provides separate magnitudes of the de Vaucouleur and
exponential components of each galaxy's light (essentially a
bulge-disk decomposition), along with the relative contribution of
each, which can be used to construct model bulge $+$ disk magnitudes and
colors for each galaxy.

The algorithm for measuring velocity dispersions has been updated in
Data Release Six (DR6) to correct for small systematic biases present
in the pre-DR6 measurements \citep{bernardi07-lsigma}.  They are
computed using the ``direct-fitting'' method \citep{burbidge61,rix92},
based on fitting stellar templates convolved with gaussian broadening
functions.  Only $\sigma$ values above the $\sim 70$ km s$^{-1}$
instrumental resolution are considered reliable.  We therefore only
use galaxies with $\log \sigma > 1.86$.  Velocity dispersion
measurement errors are typically $< 10$\% (0.04 dex), based on repeat
measurements of individual SDSS
galaxies\footnote{http://www.sdss.org/dr6/algorithms/veldisp.html}.

There is a known problem with SDSS photometry of bright galaxies
caused by scattered light that biases the sky background high.
Because of this, the total luminosities and radii of bright galaxies
are underestimated by the SDSS pipeline, as noted by \citet{lauer07},
\citet{bernardi07-bcg}, and \citet{lisker07}.  To correct for this effect,
we use the results of the simulations reported on the SDSS
website\footnote{http://www.sdss.org/dr6/start/aboutdr6.html} to
compute magnitude and radius corrections as a function of apparent
magnitude, which we apply to the galaxies in our sample.  We
additionally limit our sample to galaxies with $15 < r < 18$ to avoid
the worst of the photometry errors caused by this effect.  The
Galactic extinction corrections from the NYU Value-Added Catalog are
applied to all photometry.  

The galaxy sample presented here is limited to a modest range in
redshift ($0.04 < z < 0.08$).  This has a number of advantages.  The
SDSS Main Galaxy Sample is limited to apparent magnitude $r < 17.77$.
For $0.04 < z < 0.08$, our galaxy sample is complete down to absolute
magnitude $M_r = -20.12$, while the lower end of our redshift range
provides data down to $M_r > -19$.  The relatively small range in
redshifts means that the range in lookback times over the entire
sample is only $\Delta t \sim 0.5$ Gyr, which essentially eliminates
the effects of evolution within our sample.  At these redshifts, the
SDSS spectral fibers (1.5$''$ radius) cover a significant portion of
the galaxy, typically $\sim R_{deV}/2$.  These spectra are
therefore not nuclear spectra but sample a substantial fraction of the
galaxy light.

\begin{figure}[h]
\includegraphics[width=1.0\linewidth]{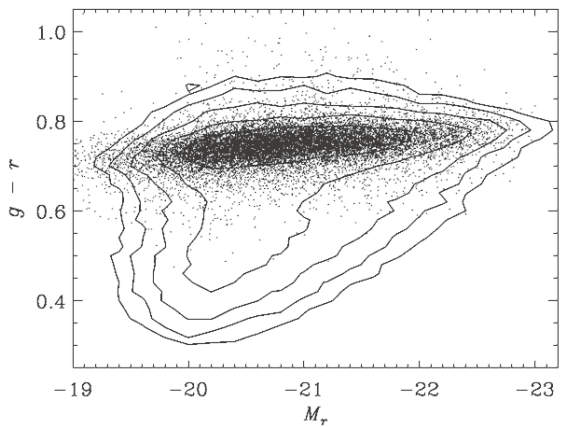}
\caption{Sample galaxies in the color-magnitude diagram.  Points show
  the galaxies that meet the sample selection criteria described in
  \S\ref{sample}, while contours show all galaxies at $0.04 < z <
  0.08$.  Galaxies in the sample are selected to be free of emission
  lines and to have light profiles characteristic of early type
  galaxies.  No explicit color selection is applied, yet the resulting
  galaxy sample consists almost exclusively of red sequence galaxies.
  The remaining color outliers are not included in the stacked
  spectra.  See \S\ref{bins} for details.  }\label{all_cmd}
\end{figure}

Throughout the rest of this paper, we will use velocity dispersion
measurements that have been corrected to $1/8$ effective radius.
Following \citet{bernardi03a} and \citet{jorgensen95}, we use $\sigma
= \sigma_{fib} (r_{fib}/\frac{1}{8}r_{\circ})^{0.04}$, where
$r_{\circ}$ is the circular galaxy radius in arc seconds, computed as
$r_{\circ} = R_{deV} \sqrt{(a/b)_{deV}}$, and $r_{fib}$ is the
spectral fiber radius, 1.5$''$ for SDSS spectra.  This allows us to
provide consistent comparisons with past stellar population work,
which typically uses central velocity dispersion measurements.  In
practice, velocity dispersion gradients are shallow, and this
correction is small (of order $\sim 6$\%).  The $r$-band and $g$-band
magnitudes used throughout this paper have been K-corrected to $z=0.0$
using the IDL code {\it kcorrect} v4.1.4 \citep{blanton03-kcorrect}
and converted to absolute magnitudes ($M_r$ and $M_g$) assuming a
standard $\Lambda$CDM cosmology with $\Lambda = 0.7$, $\Omega_M = 0.3$
and $h = 0.7$.

Spectra of the galaxies in our final sample were downloaded from the
SDSS DR6 \citep{adelman-mccarthy08} version of the SDSS
data archive server\footnote{http://das.sdss.org/DR6-cgi-bin/DAS}.  

\begin{figure*}[t]
\includegraphics[width=1.0\linewidth]{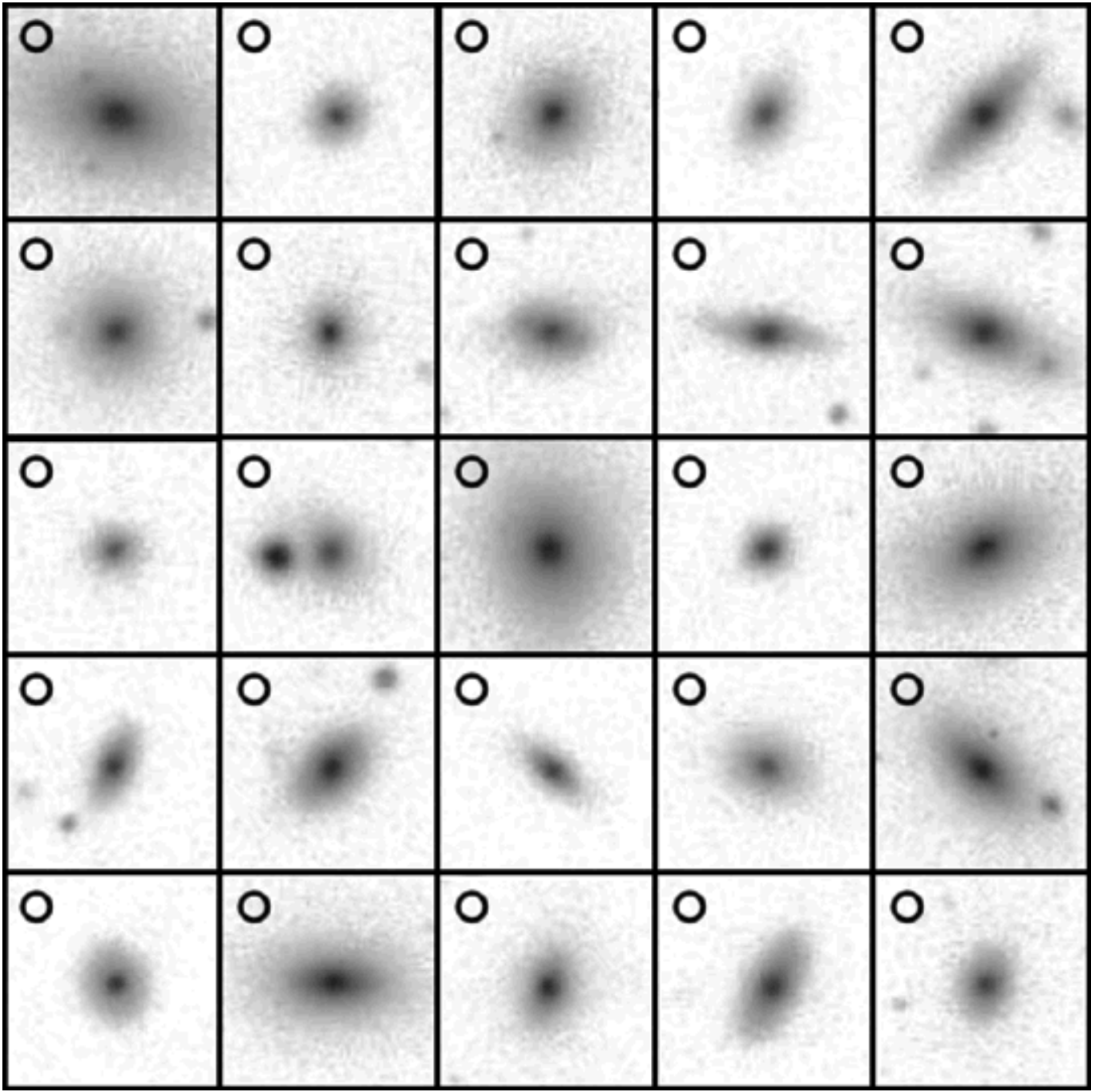}
\caption{Thumbnail images of a random selection of galaxies from the
  sample.  Circles indicate the size of the 3$''$ SDSS spectral
  fibers.  The sample is dominated by early type galaxies, but
  contains some bulge-dominated spirals as well at a range of
  orientations.  Even where a disk is present, the spectral fiber
  predominantly samples light from the bulge. }\label{galaxy_pics}
\end{figure*}

\subsection{Selection Criteria}\label{criteria}

To create a sample of passively evolving galaxies, we include only
those galaxies with no emission from ionized gas.  Specifically, we
require that the measured equivalent widths of H$\alpha$ and
[O\textsc{ii}]$\lambda$3727 emission be below a 2$\sigma$ detection
threshold, as determined by \citet{yan06}.  In this sample, the
typical 2$\sigma$ detection threshold corresponds to H$\alpha$ and
[O\textsc{ii}] equivalent widths of $\sim0.3$ {\AA} and $\sim1.7$
{\AA}, respectively.  This criterion excludes galaxies with
significant ongoing star formation within the central portion of the
galaxy covered by the spectral fiber, as well as active galactic
nuclei (AGN) with optical emission lines \citep{yan06}.  

Because the SDSS spectra only sample galaxy light with the 3$''$
spectral fiber radius, selecting quiescent galaxies in this way does
not exclude disk galaxies with quiescent central bulge populations
which harbor star formation in their outer regions.  These galaxies
are not truly quiescent galaxies, although they may have quiescent
bulges.  To limit contamination by these galaxies, we follow
\citet{bernardi03a} in excluding galaxies with $R_{90}/R_{50} < 2.5$
in the $i$-band (eliminating galaxies without centrally-concentrated
light profiles) and for which the likelihood of a de Vaucouleurs light
profile fit is less than 1.03 times the likelihood of an exponential
fit.  These further cuts reduce the sample size by $< 10$\%, leaving a
total sample of 16,008 galaxies.

In summary, the galaxies in this study have been selected from the
SDSS Main Galaxy Sample using the following criteria:
\begin{list}{}{}
\item[1.] {Redshift range: $0.04 < z < 0.08$}
\item[2.] {Apparent magnitude range: $15 < r < 18$}
\item[3.] {No detected emission in H$\alpha$ and
  [O\textsc{ii}]$\lambda3727$}
\item[4.] {Concentrated light profiles: $R_{90}/R_{50} > 2.5$ in the $i$-band}
\item[5.] {Likelihood of a de Vaucouleurs light profile $\ge 1.03$ times
  larger than the likelihoood of an exponential light profile}
\end{list}

Throughout the rest of this paper, we refer to the galaxies in our
sample as quiescent or passive galaxies, with the understanding that
this categorization refers to galaxies which are not actively forming
stars.

In Figure \ref{all_cmd}, the galaxies used in this analysis are shown
with respect to the color-magnitude distribution of all galaxies in
this redshift range.  The gray points show the galaxies included in
the sample, while the contour lines show the underlying
color-magnitude distribution of all galaxies with $0.04 < z < 0.08$
from the SDSS Main Galaxy spectroscopic sample.  In choosing the
sample presented here, no explicit color selection was applied, yet
the resulting sample of galaxies populates a narrow red sequence
relation between galaxy $M_r$ and color.  Figure \ref{galaxy_pics}
shows thumbnail images for a randomly-selected set of galaxies from
our final sample.  The galaxies all have smooth morphologies.  About
20\% of the galaxies have significant disks, although all show
evidence for prominant bulges.  In the appendix, we show that the use
of magnitudes and colors from de Vaucouleur fits to the galaxy light
profiles (even in galaxies which have a disk component) has a
negligible effect on our results.

Of our sample galaxies, 85\% have bulge fractions above 0.8 while only
3\% have bulge fractions below 0.5, making the sample galaxies highly
bulge-dominated.  Figure \ref{galaxy_pics} shows that the sample
contains a modest fraction of face-on and edge-on galaxy disks,
however these do not dominate the light of the galaxies.  Furthermore,
the disk contribution to the measured spectrum will be substantially
smaller than its contribution to the total galaxy light.  The $3''$
SDSS spectral fibers (shown as circles in Figure \ref{galaxy_pics})
sample roughly 0.3--0.6 of the typical galaxy effective radius at
these redshifts.  In face-on disks, the SDSS $3''$ spectral fiber
predominantly samples the galaxy bulge, while the spectra of inclined
disks will contain a slightly increased light contribution from the
galaxy disks.  Edge-on disks in the sample must have negligible
ongoing star formation in order to satisfy the stringent emission line
cuts.

\section{The Color-Magnitude Relation}\label{cmr}

\begin{figure*}[t]
\includegraphics[width=1.0\linewidth]{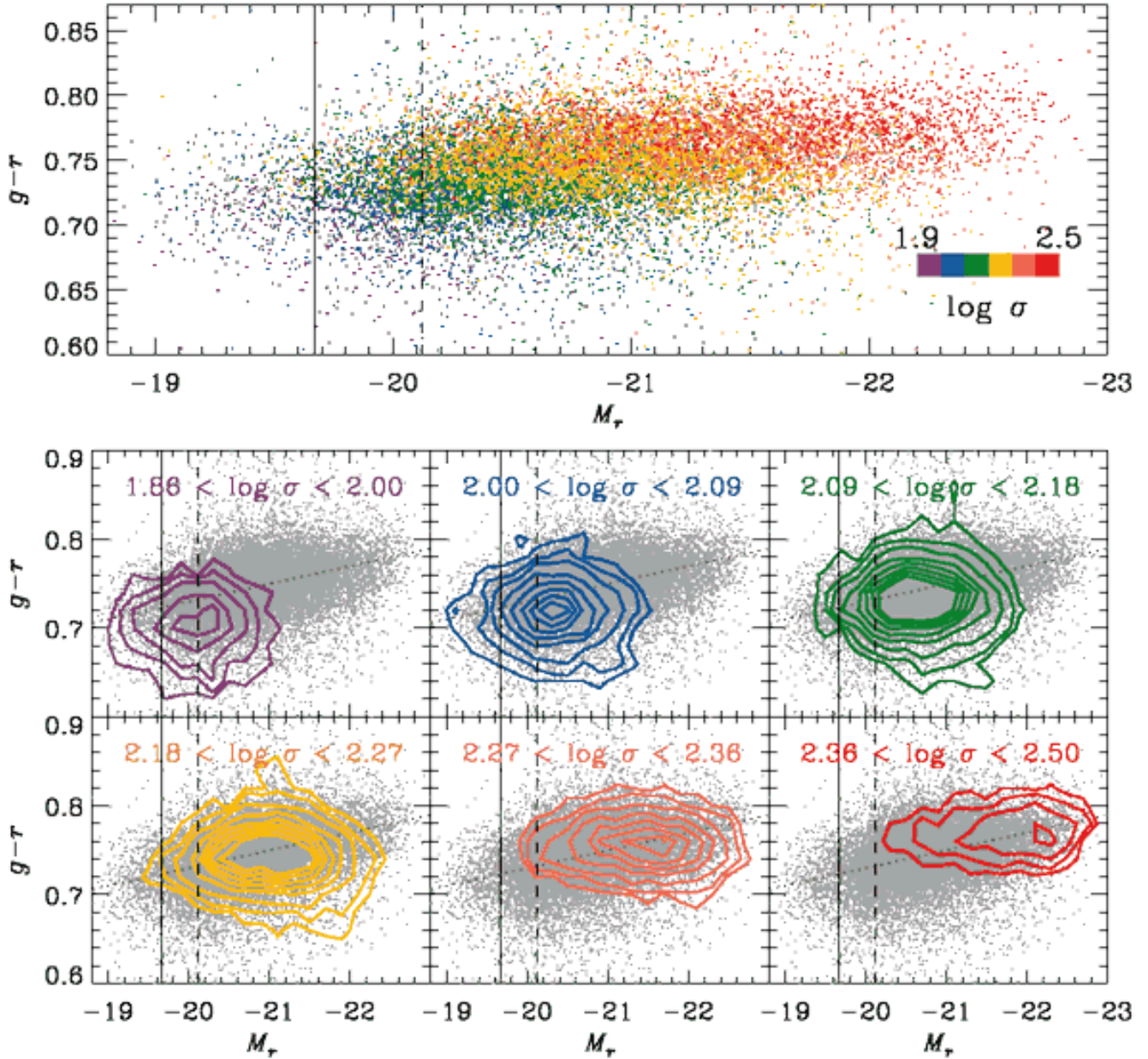}
\caption{{\it Top panel:} The color-magnitude relation for all
  galaxies in our sample.  Data points are color-coded by $\sigma$,
  with purple (red) indicating the lowest (highest) $\sigma$ bin.  The
  dashed vertical lines show the completeness limit of the sample,
  while the solid vertical lines show the magnitude at which the
  sample is missing 50\% of the galaxies (i.e., where the sample
  becomes severely incomplete).  {\it Lower panels:} The
  color--magnitude relation for galaxies at fixed $\sigma$.  Gray
  points show the color-magnitude relation for the total galaxy
  sample, while over-plotted contours show the color-magnitude
  distribution of galaxies within a limited range in $\sigma$, as
  labeled.  Contour lines indicate a number density of 5, 10, 20, 40,
  60, 80, 100, and 120 galaxies per bin, where bins in $M_r$ and $g-r$
  are defined by the tickmarks on the axes.  The gray dotted line
  indicates the color-magnitude relation for the total sample.
  Galaxies with larger values of $\sigma$ tend to be brighter and
  redder than those at lower $\sigma$.  However, galaxies at fixed
  $\sigma$ span a substantial range in both color and magnitude, so
  that different bins in $\sigma$ show a large amount of overlap in
  both color and magnitude.  At fixed $\sigma$, there is no
  color-magnitude relation.  Combining galaxies from all $\sigma$ bins
  results in an overall color-magnitude
  relation. }\label{sig_contours}
\end{figure*}

The selection criteria defined above result in a sample of quiescent
galaxies which populate the red sequence in the color-magnitude
diagram.  These galaxies show a clear color-magnitude relation (Figure
\ref{sig_contours}, top panel), such that more luminous galaxies have
redder colors.  Data points are color-coded by bins in $\sigma$ as
labelled in the lower figure panels.  In the lower panels, the total
galaxy sample is shown as a cloud of gray data points, with the total
color-magnitude relation overplotted as the gray dotted line.  In each
panel, contours indicate the color-magnitude distribution of galaxies
within a single bin in $\sigma$, as labeled.  These $\sigma$ bins are
used throughout this paper.  The vertical dashed line in Figure
\ref{sig_contours} indicates the completeness limit of the SDSS Main
Galaxy Sample in the redshift range used for this analysis.  The solid
line at $M_r = -19.7$ shows where we expect the sample is missing more
than 50\% of the galaxies at these magnitudes.  Thus the lack of
galaxies at fainter magnitudes is likely a selection effect rather
than a genuine lack of galaxies at $M_r > -19.7$.

What is notable about the contours in Figure \ref{sig_contours} is
that they show {\it no color-magnitude relation} at fixed $\sigma$, in
agreement with the work of \citet{bernardi05}.  In each panel, which
illustrates a narrow range in $\sigma$, the contours in the
color-magnitude diagram are horizontal.  However, galaxies with larger
$\sigma$ (red points) are generally more luminous and have redder
colors than those with smaller $\sigma$ (blue and purple points), so
that superposing the various $\sigma$ bins results in a net
color-magnitude relation.

\begin{figure*}[t]
\includegraphics[width=1.0\linewidth]{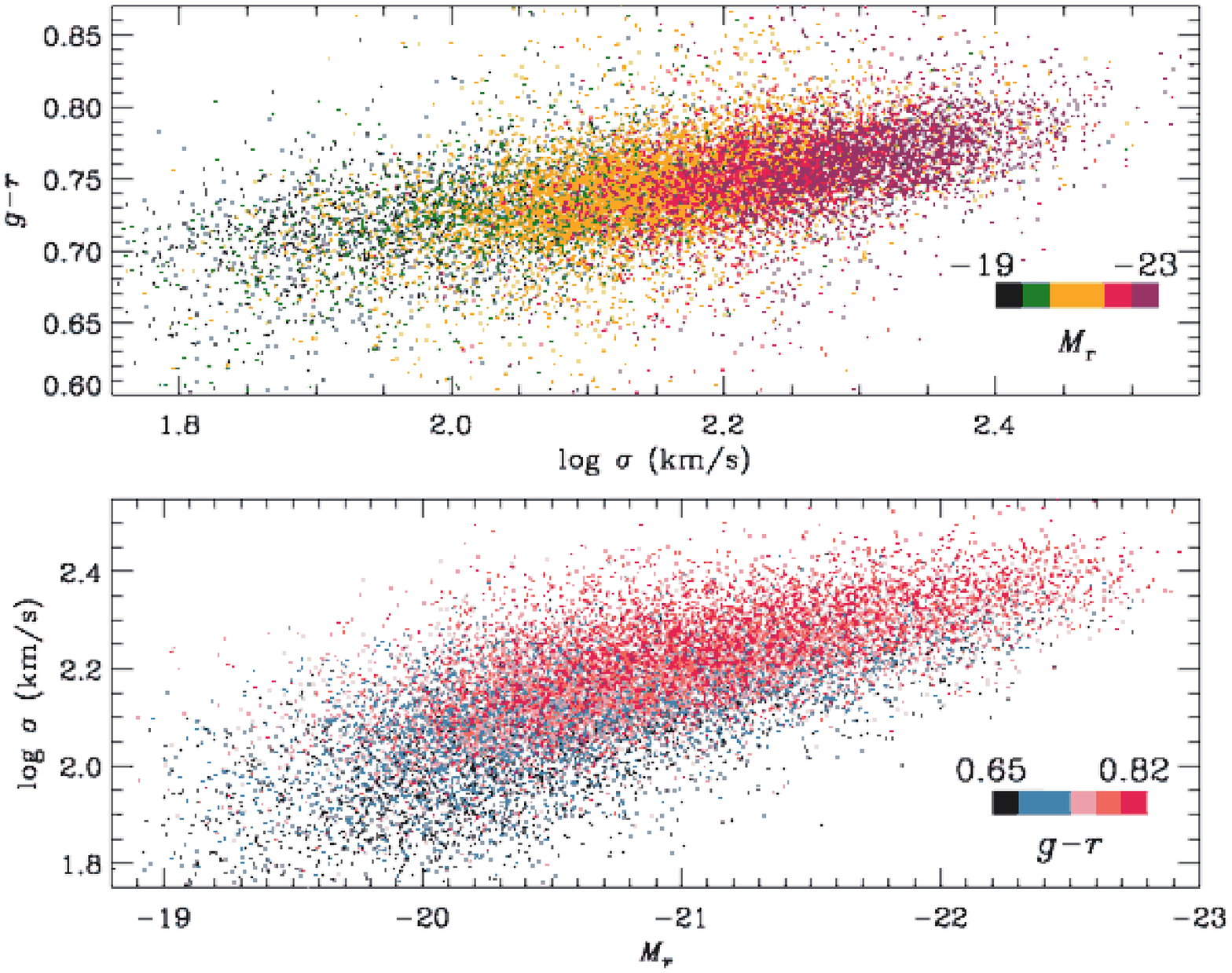}
\caption{{\it Top panel:} The color-$\sigma$ relation, with data
  points color-coded by galaxy luminosity.  Dark green (purple) points
  show the faintest (brightest) luminosity bins.  A color-$\sigma$
  relation exists at fixed $M_r$ (e.g., purple data points alone) with
  roughly the same slope as the total color-$\sigma$ relation.  The
  color-$\sigma$ trend is stronger and narrower than the
  color-magnitude relation in Figure \ref{sig_contours}.  {\it Bottom
    panel:} The $\sigma$-magnitude relation, color-coded by galaxy
  $g-r$ color.  Dark blue (dark pink) points show the bluest (reddest)
  galaxies.  The $\sigma$-magnitude relation is broader for fainter
  galaxies.  This is correlated with galaxy color, such that bluer
  galaxies tend to have lower $\sigma$ at fixed $M_r$.  This may be
  due to a greater degree of rotational support in fainter and bluer
  early type galaxies, showing lower $\sigma$ for the same $M_*$.
}\label{cmd}
\end{figure*}

In contrast, the color-$\sigma$ relation is readily apparent at fixed
$M_r$ (Figure \ref{cmd}, top).  Here, colored data points indicate
bins in luminosity, with dark green, light green, yellow, orange,
pink, and purple data points corresponding to $-18.8 > M_r > -20.0$,
$-20.0 > M_r > -20.4$, $-20.4 > M_r > -20.8$, $-20.8 > M_r > -21.2$,
$-21.2 > M_r > -21.6$, and $-21.6 > M_r > -23$, respectively.  A clear
color-$\sigma$ relation is apparent at fixed $M_r$ (e.g., purple data
points), with roughly the same slope at the total color-$\sigma$
relation.  This color-$\sigma$ correlation is stronger and tighter
than the color-magnitude relation shown in Figure \ref{sig_contours}.

Finally, the bottom panel of Figure \ref{cmd} shows the
$\sigma$-magnitude or Faber-Jackson relation, color-coded by galaxy
$g-r$ color.  Dark blue, medium blue, light blue, pale pink, medium
pink, and dark pink data points represent $g-r < 0.74$, $0.74 < g-r <
0.76$, $0.76 < g-r < 0.78$, $0.78 < g-r < 0.80$, $0.80 < g-r < 0.82$,
and $0.82 < g-r$, respectively.  The $\sigma$-magnitude relation is
tight for the most luminous galaxies, but shows considerable spread
for faint galaxies.  The spread is correlated with galaxy color, such
that bluer galaxies tend to have lower $\sigma$ at fixed luminosity.
This may indicate that fainter, bluer galaxies have a larger degree of
rotational support, resulting in lower values of $\sigma$ for a given
$M_*$.

It is clear that $\sigma$, luminosity, and color are all correlated in
early type galaxies.  \citet{bernardi05} argue that the fundamental
relations are $\sigma$-color and $\sigma$-magnitude and that the
color-magnitude relation results from a combination of these two
correlations.  The color-$\sigma$ relation (Figure \ref{cmd}, top)
exists independent of galaxy luminosity (i.e., at fixed $M_r$).
Likewise, the $\sigma$-magnitude relation (Figure \ref{cmd}, bottom)
exists independent of galaxy color.  However, the color-magnitude
relation does not exist at fixed $\sigma$ (Figure \ref{sig_contours}),
indicating that the $\sigma$ dependence of both color and magnitude is
what leads to the observed color-magnitude relation.

From Figure \ref{sig_contours}, it is also clear that galaxies with
very similar values of $\sigma$ show a substantial spread in $M_r$.
The low--$\sigma$ bins suffer from incompleteness effects, but in the
higher $\sigma$ bins (e.g., those with $\log \sigma \ge 2.18$), it is
clear that galaxies with a range of only 0.09 dex in $\log \sigma$
span roughly 2 mag (0.8 dex) in $M_r$.  Similarly, the substantial
overlap of the $\sigma$ bins in the color-magnitude diagram means that
galaxies at fixed $M_r$ have a range of $\sigma$.  

These arguments illustrate that optical luminosity $L$ is not a good
proxy for $\sigma$.  This is expected from the fact that galaxies
populate the FP and the Faber-Jackson relation is not an edge-on
projection of the FP \citep{dressler87}.  The two parameters are,
however, often used interchangeably in the literature as measures of
galaxy size or mass.  We will see in \S\ref{3v3} that using $L$ versus
using $\sigma$ as the controlling variable in stellar population
studies can lead to substantially different stellar population scaling
relations, because stellar population properties vary systematically
with $L$ at fixed $\sigma$.

\section{Dissecting the Color-Magnitude Relation}\label{deconstructing_cmr}

Accurate stellar population parameters can only be derived from high
$S/N$ spectra ($S/N \ge 100$ \AA$^{-1}$; \citealt{cardiel98}).  As
individual SDSS galaxy spectra in our sample typically have $S/N
\approx 20$ \AA$^{-1}$, dozens or more spectra must be combined to
achieve the necessary $S/N$.  If galaxies are combined randomly, the
average spectra would all look very similar.  The challenge in this
method is to identify useful ways of sorting the galaxies {\it in
advance}, in order to do the best possible job of populating the full
range of galaxy star formation histories, without averaging out all
the interesting variations.

In this section, we define a three-dimensional parameter space of
$\sigma$ (which is not sensitive to stellar population effects) along
with $L$ and color (which are) and sort galaxies into bins in this
space.  We then stack the spectra of galaxies within each bin and
measure absorption line strengths.  In \S\ref{stellar_pops}, we
combine the observed line strengths with stellar population models to
obtain the fundamental stellar population parameters of galaxies along
and across the color-magnitude relation.

\subsection{Binning by $\sigma$, $L$, and color}\label{bins}

\begin{figure}[t]
\includegraphics[width=1.0\linewidth]{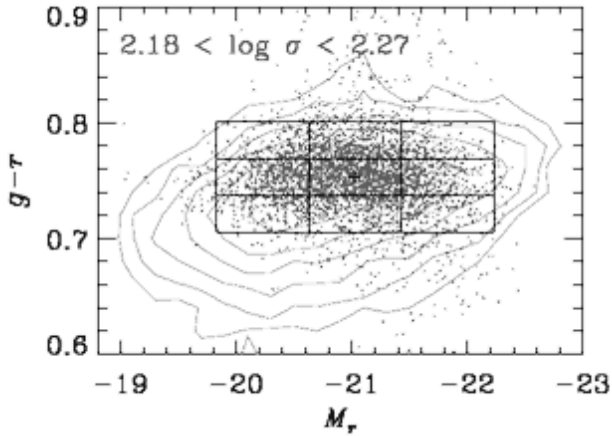}
\caption{Method used to bin galaxies by $L$ and color at fixed
  $\sigma$.  The color-magnitude relation for the total quiescent
  galaxy sample in this typical $\sigma$ bin is indicated by the gray
  contours.  Overplotted as gray points are all galaxies with $2.18 <
  \log \sigma < 2.27$.  Within this $\sigma$ bin, there is no
  color-magnitude relation.  The median point in $M_r$ and $g-r$ is
  shown as the black cross in the center.  Galaxies in this $\sigma$
  bin are divided into a 3x3 grid in color-magnitude space, with the
  grid centered on the median point and with fixed grid spacing of 0.8
  mag in $M_r$ and 0.032 mag in $g-r$.  Defining the grid in this way
  eliminates color and $L$ outliers which may bias the average
  properties of the bins.  Galaxies in each of the six $\sigma$ bins
  are sorted in this way into nine sub-bins in color--$L$ space, and
  the galaxies in each sub-bin are stacked as described in
  \S\ref{stacking} to produce high $S/N$ mean spectra for all 54 bins
  in $\sigma$--$L$--color space.}\label{sig4_bins}
\end{figure}

In \S\ref{cmr}, we divided our sample of quiescent galaxies into six
bins based on $\sigma$.  To study how stellar populations vary at
fixed $\sigma$, we further divide those same $\sigma$ bins into three
bins in color by three bins in $L$ for a total of 54 bins.  Figure
\ref{sig4_bins} illustrates this process for the $\log \sigma =
2.18$--2.27 bin.  Cuts in color and $L$ are defined with fixed width,
such that each $L$ bin is 0.8 mag wide in $M_r$ and each color bin is
0.032 mag wide in $g-r$.  The 3x3 grid of sub-bins is centered on the
median $M_r$ and $g-r$ for all the galaxies in the given $\sigma$ bin.
The median values of $\sigma$, $M_r$, and $g-r$ for each of the 54
bins are listed in Table \ref{bin_tab}, along with the total number of
galaxies in each bin.

The central sub-bin in color--$L$ space contains more galaxies than
the outer bins, but all bins are sufficiently well populated to
produce high $S/N$ stacked spectra.  Defining the color--$L$ grid in this
way excludes color and $L$ outliers from each $\sigma$ bin, which
ensures that the stacked spectra will not be biased by very deviant or
misidentified objects.

In the lowest $\sigma$ bins, the magnitude limit of the sample
presented here means that the fainter galaxies at that $\sigma$ are
under-represented.  Under the assumption that the missing faint
galaxies at $z \sim 0.08$ are not substantially different from those
at $z \sim 0.04$ which are included in the sample, the fainter bins
are not strongly biased but are less populated than they ought to be.
Comparisons between galaxies at different $L$ in the same $\sigma$ bin
should therefore be robust to completeness effects.  However, because
these $\sigma$ bins are missing the faintest galaxies, the median
$M_r$ used to locate the color--$L$ grid may be biased brighter than
the underlying galaxy population at that $\sigma$.

\subsection{Constructing the stacked spectra}\label{stacking}

In each of the 54 bins in $\sigma$--$L$--color space, we combine
spectra of the galaxies in the bin in order to produce a very high
$S/N$ average spectrum.  Regions around the bright skylines at
5577{\AA}, 6300{\AA}, and 6363{\AA} are masked in the individual
galaxy spectra, which are then shifted to the restframe.  The
individual spectra are normalized so that all galaxies in the bin
contribute equally to the averaged spectrum.  The normalization uses
the median flux in the 4100--5000{\AA} range without modifying the
continuum shape, as the flux-calibrated spectral shape affects the
Lick index measurements.  In practice, because the galaxies in a given
bin have very similar colors, the shape of the continuum is nearly
identical for all galaxies in the bin.

The stellar absorption features in higher $\sigma$ galaxies are
effectively observed at lower resolution than those in lower $\sigma$
galaxies, due to the increased intrinsic Doppler smoothing of the
lines from stellar motions within the galaxy.  This intrinsic
smoothing has non-negligible effects on the measured absorption line
strengths.  In order to compare all galaxies at the same effective
resolution, we smooth lower--$\sigma$ galaxies up to match the
highest-$\sigma$ galaxies in our sample, $\sigma \approx 300$ km
s$^{-1}$.  The algorithm we use to stack the galaxy spectra first
averages the unsmoothed spectra of galaxies within narrow $\sigma$
sub-bins, with a routine that rejects very deviant pixels.  The
average clean sub-bin spectra are then smoothed to the maximum
$\sigma$ of the sample ($\sim 300$ km s$^{-1}$) and coadded.  The
result is a stacked spectrum for each $\sigma$--$L$--color bin from
which outlier pixels have been rejected and which has been smoothed to
an effective resolution identical to all the other bins.  The error
spectrum is also computed from the individual SDSS galaxy error
spectra by adding the individual pixel errors in quadrature to produce
an error spectrum for each stacked spectrum.

\subsection{Lick Index absorption strengths}\label{indices}

We measure the full set of Lick indices in each of the stacked spectra.  These
include the Balmer lines H$\beta$, H$\gamma_F$, and H$\delta_F$ (also
broad definitions of H$\gamma_A$ and H$\delta_A$), and a set of
Fe-dominated lines (Fe4383, Fe4531, Fe5015, Fe5270, Fe5335, Fe5406,
Fe5709, and Fe5782), as well as numerous lines that are sensitive to
abundances of elements other than Fe (Mg$_1$, Mg$_2$, Mg~{\it b},
CN$_1$, CN$_2$, C$_2$4668, Ca4227, Ca4455, NaD, TiO$_1$, and TiO$_2$).
The index definitions are taken from \citet{worthey94} and
\citet{worthey97}, and line strengths are measured using the {\it
  Lick\_EW} code that is available
online\footnote{http://www.ucolick.org/$\sim$graves/EZ\_Ages.html} as
part of the {\it EZ\_Ages} code package \citep{graves08}.  The {\it
  Lick\_EW} code also computes statistical errors for each Lick index
based on the error spectra, using equations 33 and 37 of
\citet{cardiel98}.  Because the Lick indices are defined as equivalent
widths, they are insensitive to dust extinction effects.  

All the stacked spectra have been smoothed to $\sigma = 300$ km
s$^{-1}$.  Combined with the intrinsic resolution of the SDSS
spectrograph, the spectra are at {\it lower} resolution than the
Lick/IDS resolution at which the Lick indices are defined.  A
correction must be applied to the measured line strengths to bring
them back onto the Lick index system.  These corrections are included
in the {\it Lick\_EW} computation, based upon smooth single stellar
population spectra, as given in \citet{schiavon07} Table A2a.  All of
the spectra have been smoothed to the same $\sigma$ and corrected in
the same way so that uncertainties in the velocity dispersion
correction should not affect relative lines strength measurements.
The corrected Lick index measurements and statistical errors for each
of the 54 stacked spectra are given in Table \ref{index_tab}.  We do
not attempt to match the zeropoints of the SDSS spectra to those of
the Lick system as defined by \citet{schiavon07}.  This should only
introduce minor zero-point uncertainties because the Schiavon models
are based on flux-calibrated spectra and the SDSS spectra are also
flux calibrated (see \citealt{schiavon07}, section 2.2.2).

\subsection{Stellar population modelling}\label{model}

We compare the Lick index measurements for each of the stacked spectra
to the stellar population models of \citet{schiavon07} to derive
fundamental stellar population parameters from the line strengths.
The \citet{schiavon07} models are single burst models which include
the effect of variable abudance ratios by combining theoretical
stellar isochrones \citep{girardi00} with a library of empirical
stellar spectra \citep{jones99} and individual line strength
sensitivities to elemental abundance variations computed from
theoretical stellar atmospheres \citep{korn05}.  With these models, a
set of Lick index measurements can be used to determine the mean
luminosity-weighted stellar population age, iron abundance ([Fe/H]),
and abundance ratios for the elements Mg, C, N, and Ca ([Mg/Fe],
[C/Fe], [N/Fe], and [Ca/Fe]) using the algorithm described in
\citet{graves08} and implemented in the publicly
available\footnote{http://www.ucolick.org/$\sim$graves/EZ\_Ages.html}
code {\it EZ\_Ages}.  Briefly, the \citet{graves08} method determines
a fiducial mean age and [Fe/H] from a combination of Balmer and
Fe-dominated indices, then uses other indices which are sensitive to
Mg, C, N, and Ca to adjust the element abundance ratios until the
model index predictions match the ensemble of Lick index data.  The
basic modelling process has not changed since the original analysis of
red sequence galaxies with LINER-like emission in \citet{graves07};
subsequent minor updates to the {\it EZ\_Ages} code have improved the
robustness of the code at runtime and produce results that match those
of \citet{graves07} within the quoted errors.  {\it EZ\_Ages}
estimates stastical errors for each of the stellar population
parameters, based on the measurements errors in the Lick indices.  The
errors in these various parameters are not independent; an analysis of
the correlated errors is presented in Figure 3 of \citet{graves08}.
The reader is referred to that article for additional details on the
age and abundance fitting process.

\citet{graves08} have shown that this fitting method is robust to the
choice of Lick indices used in the analysis.  Here we use the
``standard set'' of Lick indices from \citet{graves08}---H$\beta$,
$\langle$Fe$\rangle$ (which is an average of the Fe5270 and Fe5335
indices), Mg~{\it b}, C$_2$4668, CN$_1$, and Ca4227---with the {\it
EZ\_Ages} code to fit a mean luminosity-weighted age, [Fe/H], [Mg/Fe],
[C/Fe], [N/Fe], and [Ca/Fe] for each of the 54 average galaxy spectra.
We can thus track each of these stellar population parameters as a
function of $\sigma$, $L$, and color for the galaxies in our sample.

\begin{figure}[t]
\includegraphics[width=1.0\linewidth]{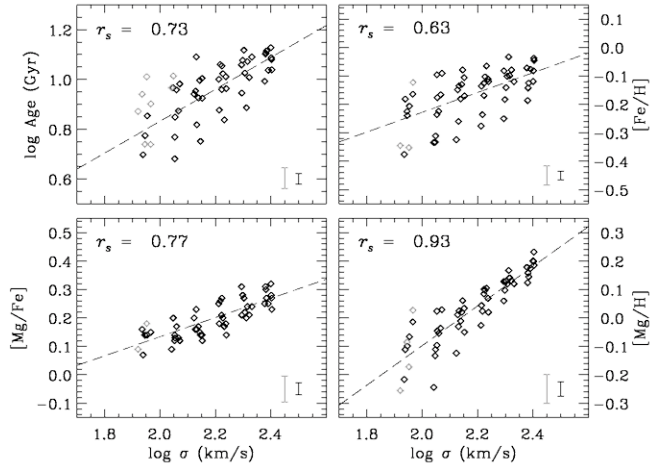}
\caption{Mean stellar population parameters for each of the 54 stacked
  galaxy spectra, shown as a function of median $\sigma$ for the
  galaxies in each bin.  The error bars in the lower right corner of
  each panel show the median statistical error in the derived
  population parameters, based on the uncertainties in the Lick index
  measurements.  The data have been separated into those with small
  statistical errors (black) and those with larger statistical errors
  (gray), as indicated by the corresponding error bars.  ``Large''
  statistical errors are defined as those more than two standard
  deviations larger than the median error.  Mean luminosity-weighted
  stellar age, [Fe/H], [Mg/H], and [Mg/Fe] all increase with
  increasing $\sigma$.  Dashed lines show linear least squares fits of
  the stellar population properties onto $\log \sigma$, computed using
  the black data points only.  The Spearman rank correlation
  coeffecients ($r_s$) of the correlations are specified in each
  panel.  The scatter in mean age and [Fe/H] is significantly larger
  than that expected due to measurement errors, implying genuine
  variation in these population parameters at fixed $\sigma$.  The
  spread in age is small for the highest--$\sigma$ galaxies, all of
  which show old ages, while the lowest--$\sigma$ galaxies cover a
  substantial factor in mean age.  There is only modest scatter in
  either [Mg/Fe] or [Mg/H] at fixed $\sigma$.  The correlation bewteen
  $\sigma$ and [Mg/H] is particulary strong, as indicated by the very
  high value of $r_s$.  }\label{sig_v4}
\end{figure}

In this analysis, we focus on the stellar population age, [Fe/H],
[Mg/H], and [Mg/Fe], leaving the analysis of other abundance ratios
for future work.  The results of the stellar population analysis for
the stacked spectra are presented in Table \ref{index_tab}.  Of the
$\alpha$-elements, Mg is the only one that is relatively
straightfoward to measure in optical galaxy spectra.  For the purposes
of this type of stellar population work, it is typically assumed that
all $\alpha$-elements scale together and thus that [Mg/Fe] is
equivalent to [$\alpha$/Fe].  There is some observational evidence
that $\alpha$ elements may not always vary in lock-step (e.g.,
\citealt{fulbright07,humphrey06}) but, lacking an alternative, we will
use [Mg/Fe] as an estimate of the total [$\alpha$/Fe].  The
\citet{schiavon07} models are computed at fixed [Fe/H] rather than at
fixed total metallicity ([Z/H]), so that enhancements in individual
elements also change the total metallicity of the galaxy.

It is important to keep in mind that the stellar population parameters
based on modelling absorption line indices give ages and abundances
averaged over the ensemble of stars in the galaxy.  The contribution
of each star to the average age or abundance measurement is
proportional to the light contributed at the wavelength of the
absorption feature in question, and also to the equivalent width of
the particular absorption feature of that star.  When comparing ages
of galaxies, a younger mean luminosity-weighted age in one galaxy
compared to another does not necessarily mean that all of the stars in
the galaxy are younger: a galaxy may contain a fraction of young stars
that skew the average age to lower values.  The luminosity-weighted
ages presented here are also averaged over the ensemble of galaxies in
the bin and should thus be treated as a statistical description of
galaxies in that bin rather than as ``true'' ages for all of the stars
in an individual galaxy.

\begin{figure}[t]
\includegraphics[width=1.0\linewidth]{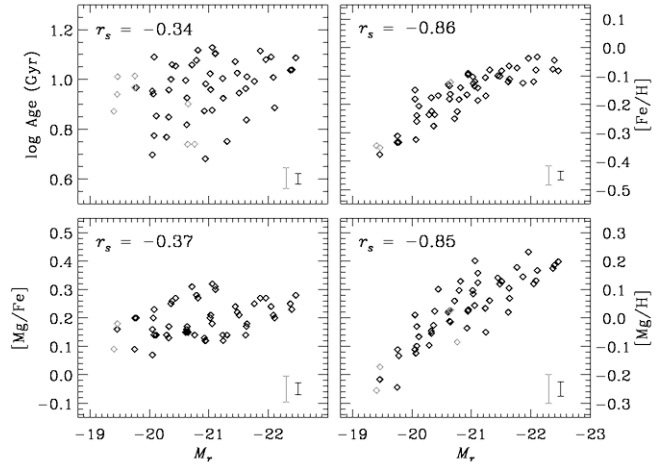}
\caption{Mean stellar population parameters as a function of galaxy
  $L$.  The results of the stellar population analysis are plotted
  against the median $M_r$ for the galaxies in each bin.  Statistical
  errors in the derived population parameters are shown in the lower
  right corner of each panel.  The data have been separated into those
  with small statistical errors (black) and those with larger
  statistical errors (gray), as indicated by the corresponding error
  bars.  The Spearman rank correlation coefficient ($r_s$) is given
  for each relation.  The strong increases in mean age and [Mg/Fe]
  visible in Figure \ref{sig_v4} when plotted against $\sigma$ are
  considerably washed out when plotted against $M_r$, while the trend
  between [Fe/H] and $M_r$ is much tighter than the $\sigma$--[Fe/H]
  relation.  The [Mg/H]--$M_r$ is still strong, although $r_s$
  indicates it is slightly weaker than the $\sigma$--[Mg/H]
  correlation.}\label{lum_v4}
\end{figure}

\begin{figure}[t]
\includegraphics[width=1.0\linewidth]{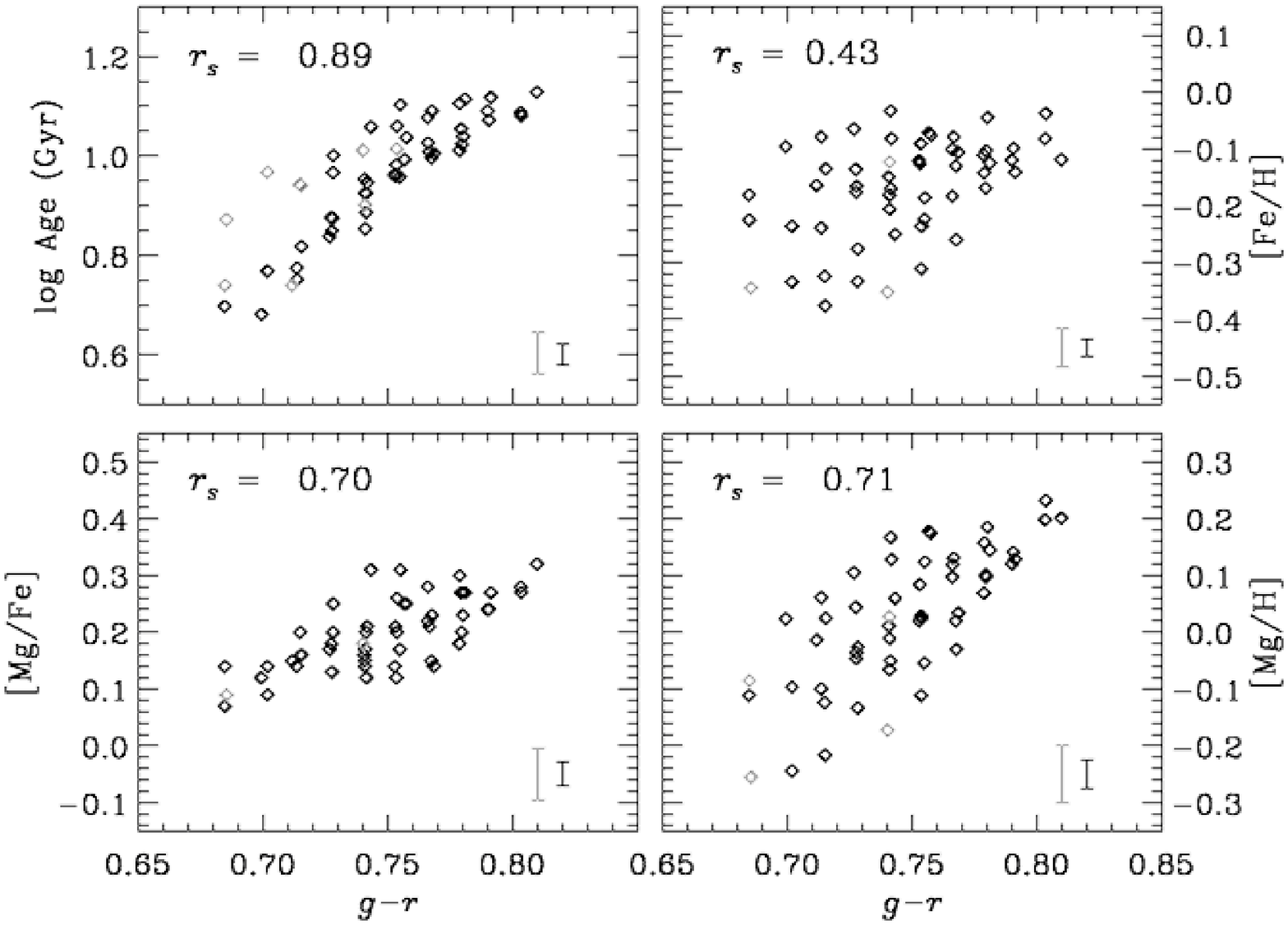}
\caption{Mean stellar population parameters as a function of galaxy
  color.  The results of the stellar population analysis are plotted
  against the median $g-r$ color for the galaxies in each bin.
  Statistical errors in the derived population parameters are shown in
  the lower right corner of each panel.  The data have been separated
  into those with small statistical errors (black) and those with
  larger statistical errors (gray), as indicated by the corresponding
  error bars.  The Spearman rank correlation coefficienct ($r_s$) is
  given for each relation.  Here, mean age increases with color,
  similar to the $\sigma$--age trend observed in Figure \ref{sig_v4}
  although with somewhat smaller scatter in log age at fixed $g-r$
  than was seen at fixed $\sigma$.  However, [Fe/H] shows only a very
  weak trend with color, while [Mg/Fe] and [Mg/H] increase
  significantly in redder galaxies.  Neither the [Mg/Fe]--color nor
  [Mg/H]--color trends are as strong as the those with $\sigma$, as
  indicated by $r_s$.}\label{color_v4}
\end{figure}

\section{Stellar Populations on the Red Sequence}\label{stellar_pops}

\subsection{Age, [Fe/H], [Mg/H], and [Mg/Fe] as Functions of $\sigma$, $L$,
  and Color}\label{3v3}

Figure \ref{sig_v4} shows the age, [Fe/H], [Mg/H], and [Mg/Fe] results
of the stellar population analysis as a function of galaxy $\sigma$.
The stellar population parameters derived from the stacked spectra are
plotted against the median value of $\sigma$ for the ensemble of
galaxies in each of the 54 bins.  The median statistical errors in the
derived population parameters are shown in the lower right corner of
each panel.  The Spearman rank correlation coefficient
($r_s$)\footnote{For a given pair of parameters, $r_s$ indicates the
  extent to which they produce the same rank ordering of objects, with
  0 indicating no correlation and 1 indicating perfect correlation.
  The sign of $r_s$ indicates whether the correlation is positive or
  negative.  The definition of $r_s$ makes no assumptions about the
  functional form of the correlation and allows robust comparison of
  correlation strengths between different correlations.}  is shown for
each pair of parameters, indicating the strength of the correlation.
Age, [Fe/H], [Mg/H], and [Mg/Fe] all show correlations with $\sigma$,
such that higher--$\sigma$ galaxies have older mean ages, higher iron
abundances, higher magnesium abundances, and higher levels of
Mg-enhancement, which strongly suggests that higher--$\sigma$ galaxies
also have higher total metallicities.  These results are in agreement
with numerous previous authors (e.g., \citealt{bernardi03d};
\citealt{thomas05}; \citealt{nelan05}; \citealt{smith07}; see summary
in Table 8 of \citealt{graves07}), although these trends are not
universally observed in all early type galaxy samples (e.g.,
\citealt{kelson06}; \citealt{trager08}).

The [Mg/H]-$\sigma$ and [Mg/Fe]-$\sigma$ relations are particularly
tight compared to the typical measurement errors.  This implies that
at fixed $\sigma$ [Mg/H] and [Mg/Fe] vary only weakly with the other
two global parameters studied here ($L$ and color).  At high $\sigma$,
[Mg/H] shows almost no variation at fixed $\sigma$. A linear
least-squares fit of [Mg/Fe] onto $\sigma$ gives a slope of 0.34,
which is consistent with previous results.

\begin{figure*}[t]
\includegraphics[width=1.0\linewidth]{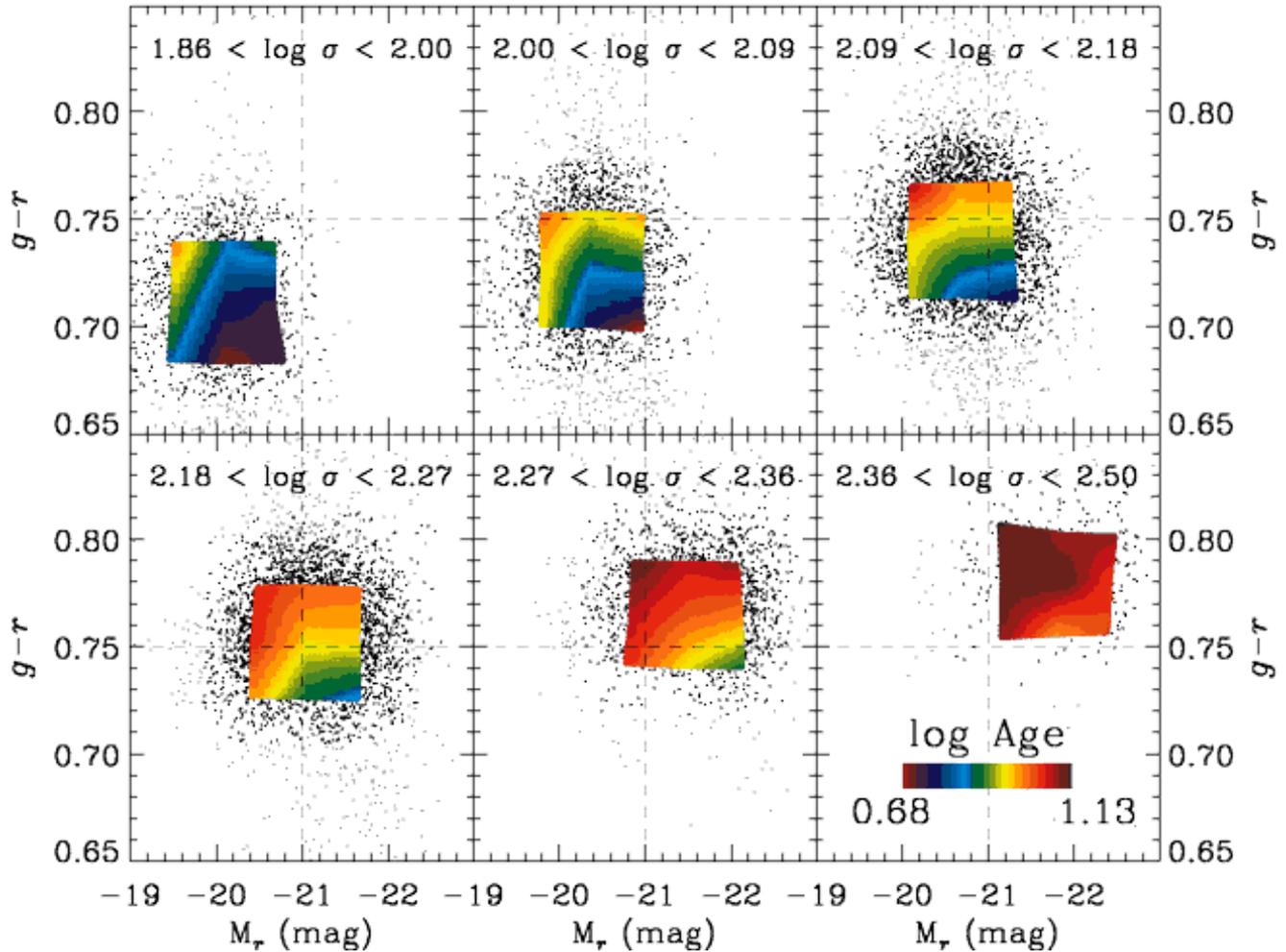}
\caption{Mean luminosity-weighted age as a function of $\sigma$, $L$,
  and color.  Each panel shows the color-magnitude diagram for a slice
  in $\sigma$, corresponding to the six $\sigma$ bins in Figure
  \ref{sig_contours}.  Black points represent galaxies that fall
  within 1.2 mag of the median $M_r$ and within 0.048 mag of the
  median $g-r$ for the $\sigma$ bin.  Gray points show galaxies that
  fall outside these boundaries and which are excluded from the
  stacked spectra, as described in \S\ref{bins}.  Overplotted are
  contours of mean luminosity-weighted galaxy age, derived from the
  stellar population modelling in \S\ref{model}.  The dashed lines are
  reference lines to guide the eye when comparing between $\sigma$
  bins.  Stellar population age increases with increasing $\sigma$,
  but shows substantial scatter within a fixed $\sigma$ slice.  There
  is more age variation amongst low--$\sigma$ galaxies than in
  high--$\sigma$ galaxies.  At fixed $\sigma$, lines of constant age
  run roughly diagonally across the color-magnitude diagram, with
  younger galaxies having bluer colors and brighter luminosities,
  while older galaxies have redder colors and fainter luminosities.
  When the various $\sigma$ slices are stacked together into the
  observed red sequence, the decrease in age with $L$ at fixed
  $\sigma$ counter-acts the overall increase of age with $\sigma$,
  erasing the age trend when $L$ is used to parameterize galaxies
  instead of $\sigma$ (as in Figure \ref{lum_v4}).  See
  \S\ref{mapping} for details.  }\label{cmd_age}
\end{figure*}

Unlike [Mg/H] and [Mg/Fe], both age and [Fe/H] show scatter at fixed
$\sigma$ that is large compared to the statistical error bars.  For
[Fe/H], the scatter is roughly the same at all $\sigma$, whereas the
spread in mean ages at fixed $\sigma$ appears to be larger for
lower--$\sigma$ galaxies.  It is noteworthy that some low--$\sigma$
galaxies have mean ages nearly as old as some of the highest-$\sigma$
galaxy bins.  Thus any model for galaxy formation that predicts a
correlation between galaxy $\sigma$ and galaxy age must also be able
to accommodate low--$\sigma$ galaxies with old ages.

Because $\sigma$ correlates with both $L$ and color (see \S\ref{cmr}),
one might expect to see similar correlations between stellar
population properties and $L$ or color as were seen for $\sigma$.
Figure \ref{lum_v4} shows the derived stellar population parameters as
functions of $L$, while Figure \ref{color_v4} shows the same as
functions of galaxy color.  Interestingly, when the stellar population
properties are plotted against $L$ instead of $\sigma$, the
correlations with age and [Mg/Fe] are nearly erased, while the
correlation with [Fe/H] is much tighter.  Conversely, when they are
plotted against color, the correlation with mean stellar age is
slightly tighter than with $\sigma$, while the [Fe/H] and [Mg/Fe]
correlations are weaker than with $\sigma$.  Of the four stellar
population parameters, [Mg/H] alone shows relatively strong
correlations with all three global parameters, although it is most
strongly correlated with $\sigma$.

Figures \ref{sig_v4}--\ref{color_v4} show that, although $\sigma$,
$L$, and color are all correlated with one another, the various
stellar population parameters show significantly different behavior
with each of the global parameters.  Residuals from the
$\sigma$--stellar population relations must correlate with $L$ and
color in such a way as to erase some trends and strengthen others when
$L$ or color are used instead of $\sigma$.  This correlation of
residuals immediately suggests that quiescent galaxies populate a
multi-dimensional stellar population parameter space.  To explore
these effects in detail, we must investigate the dependence of stellar
age, [Fe/H], [Mg/H], and [Mg/Fe] on all three global parameters
simultaneously.

\begin{figure*}[t]
\includegraphics[width=1.0\linewidth]{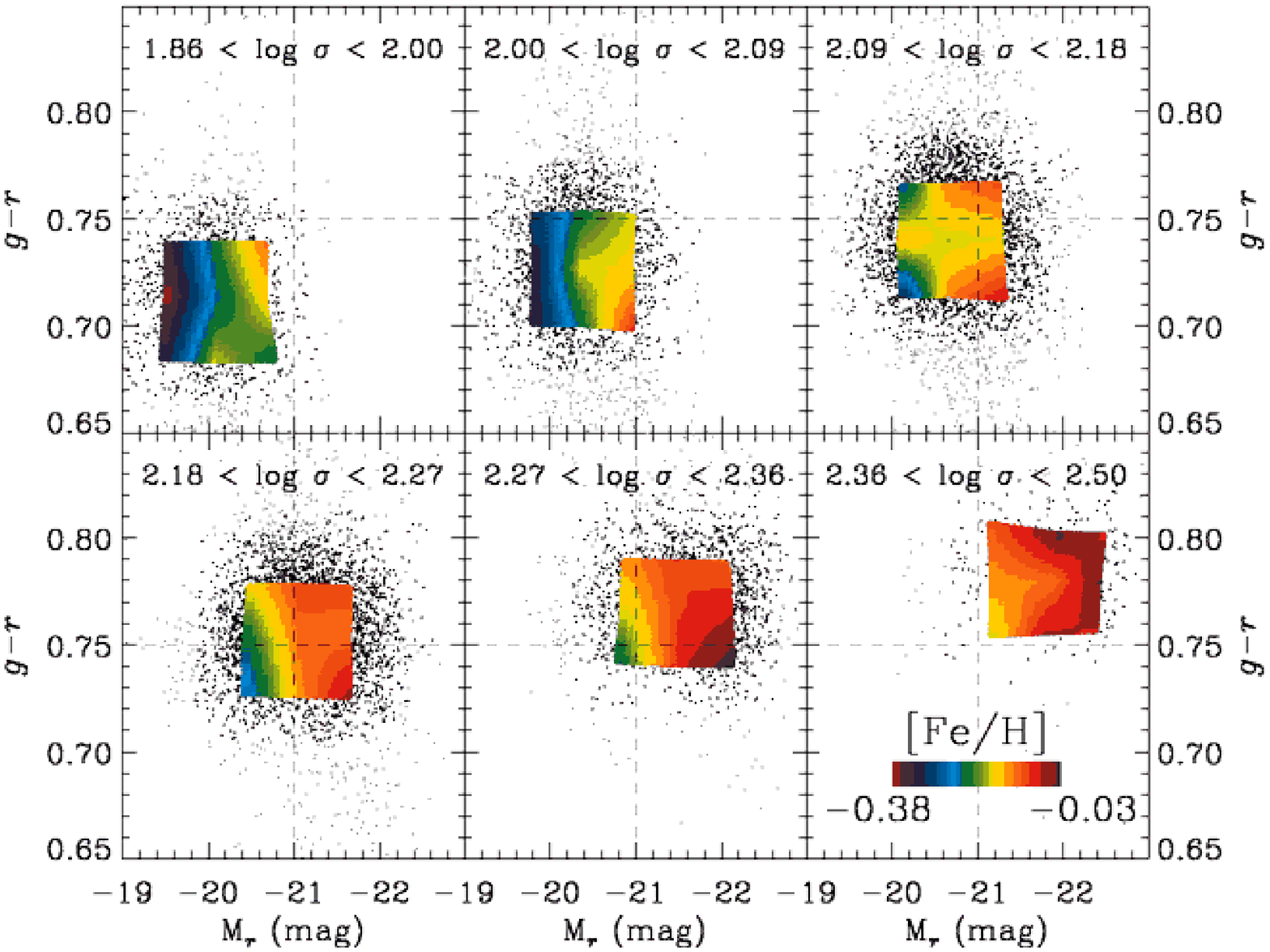}
\caption{[Fe/H] as a function of $\sigma$, $L$, and color.  As in
  Figure \ref{cmd_age}, panels show the color-magnitude diagram for
  slices of fixed $\sigma$.  Black points show galaxies included in
  the stacked spectra, while gray points show color and $M_r$ outliers
  that have been excluded from this analysis.  Overplotted are
  contours of [Fe/H] derived from the stellar population modelling of
  \S\ref{model}.  [Fe/H] increases with increasing $\sigma$.  Lines of
  constant [Fe/H] run approximately vertically in color-magnitude
  space at fixed $\sigma$, with the most luminous galaxies being the
  most Fe-rich.  When the various $\sigma$ slices are stacked together
  into the observed color-magnitude relation, the increase in [Fe/H]
  with $L$ at fixed $\sigma$ reinforces the increase in [Fe/H] with
  $\sigma$, causing the $L$--[Fe/H] relation to be tighter and steeper
  than the $\sigma$--[Fe/H] relation.  Comparisons with Figure
  \ref{cmd_age} show that, at fixed $\sigma$, age and [Fe/H] are
  roughly anti-correlated, helping to preserve the tightness of the
  total color--$\sigma$ relation.  }\label{cmd_feh}
\end{figure*}

\subsection{Mapping Stellar Populations Across the Red
  Sequence}\label{mapping}

Figure \ref{cmd_age} shows contours of stellar population
age as a function of $\sigma$, $L$, and color.  The six panels contain
data for our six standard $\sigma$ slices corresponding to the panels
in Figure \ref{sig_contours}.  Black and gray data points in each
panel show the color-magnitude relation at that $\sigma$, with gray
points indicating galaxies in the $\sigma$ range which were excluded
from the stacked spectra.

Plotted over the color-magnitude relations in each bin are color
contours representing the mean luminosity-weighted ages derived from
the stacked spectra.  The age contours were constructed as follows:
age values from the stacked spectra in each $\sigma$ range were
plotted at the median values of $M_r$ and $g-r$ for each bin, forming
a 3x3 grid of age values in color-magnitude space.  Values of age were
linearly interpolated (in log age) between the nine grid points to
produce age contours across the color-magnitude relation.  The color
bar in the lower right panel indicates the scaling of the age
contours, with the lowest and highest values indicated.  The dashed
lines at $M_r = -21$ and $g-r = 0.75$ are the same in all panels and
exist merely to guide the eye.  As $\sigma$ increases, galaxies become
systematically both brighter and redder with respect to the dashed
lines.

The age contours in Figure \ref{cmd_age} illustrate the same trends
that were visible in Figure \ref{sig_v4}.  As $\sigma$ increases from
low (upper left panel) to high values (lower right panel), the average
age indicated by the contours increases.  However, in each slice in
$\sigma$, a range of ages exists.  The spread in ages at fixed
$\sigma$ is large in the low--$\sigma$ bins and becomes increasingly
smaller toward the highest--$\sigma$ bins.

\begin{figure*}[t]
\includegraphics[width=1.0\linewidth]{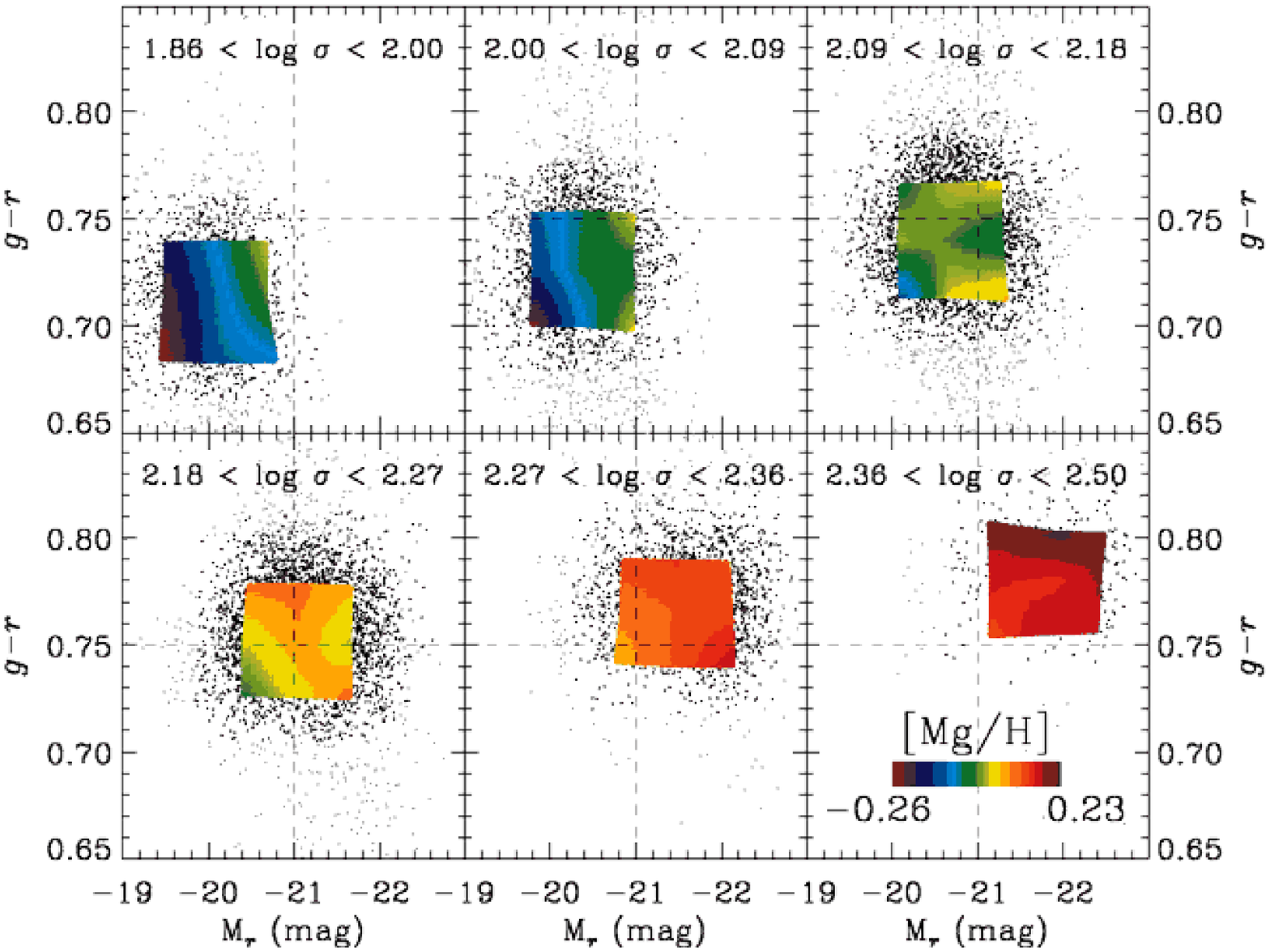}
\caption{[Mg/H] as a function of $\sigma$, $L$, and color.  Panels
  show the color-magnitude diagram for slices of fixed $\sigma$.
  Black points shows galaxies included in the stacked spectra, while
  gray points show color and $M_r$ outliers that have been excluded
  from this analysis.  Overplotted are contours of [Mg/H] derived from
  the stellar population modelling of \S\ref{model}.  Mg is used to
  trace $\alpha$-elements, thus contours of [Mg/H] here are a proxy
  for contours of [$\alpha$/H].  [Mg/H] increases with increasing
  $\sigma$, showing the least variation at fixed $\sigma$ of any of
  the stellar population parameters.  In the lower--$\sigma$ bins,
  [Mg/H] behaves similarly to [Fe/H], with lines of constant [Mg/H]
  running nearly vertically in color--magnitude space.
}\label{cmd_mgh}
\end{figure*}

The contours also illustrate the behavior of stellar population age at
fixed $\sigma$ {\it across} the color-magnitude relation.  Lines of
constant age thus run roughly diagonally in color-magnitude space at
fixed $\sigma$, with young ages in the lower right corner (bright and
blue) and old ages in the upper left corner (faint and red).  This
trend is approximately repeated in each $\sigma$ slice, although at
high $\sigma$ the dynamic range is less.  The variation at fixed
$\sigma$ cannot be explained by errors in $\sigma$ that result in
galaxies being assigned to the wrong stacked spectrum because this
would result in the opposite effect: galaxy contamination from
higher--$\sigma$ bins would tend to make the more luminous galaxies
appear older, not younger.

\begin{figure*}[t]
\includegraphics[width=1.0\linewidth]{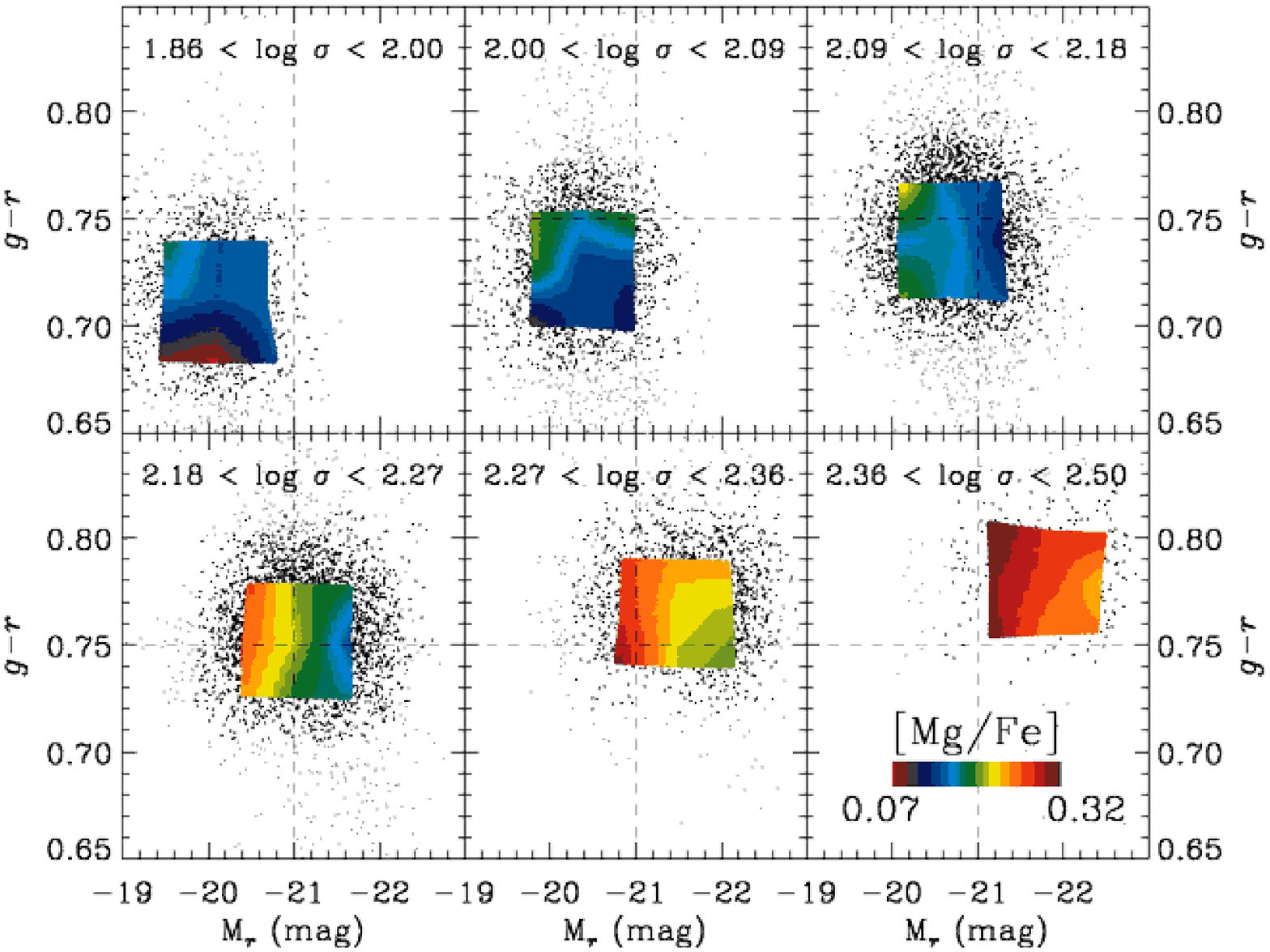}
\caption{[Mg/Fe] as a function of $\sigma$, $L$, and color.  Panels
  show the color-magnitude diagram for slices of fixed $\sigma$.
  Black points shows galaxies included in the stacked spectra, while
  gray points show color and $M_r$ outliers that have been excluded
  from this analysis.  Overplotted are contours of [Mg/Fe] derived
  from the stellar population modelling of \S\ref{model}.  Mg is used
  to trace $\alpha$-elements, thus contours of [Mg/Fe] here are a
  proxy for contours of [$\alpha$/Fe].  [Mg/Fe] increases with
  increasing $\sigma$.  There is limited variation in [Mg/Fe] at fixed
  $\sigma$.  A comparisons with Figure \ref{cmd_age} shows that
  [Mg/Fe] varies similarly to age, such that fainter galaxies at fixed
  $\sigma$ are older and have higher [Mg/Fe].  This variation runs in
  the opposite sense of variation in [Fe/H] at fixed $\sigma$, which
  increases with increasing luminosity.  When the various $\sigma$
  slices are stacked together in the observed red sequence, the
  decrease of [Mg/Fe] with increasing $L$ at fixed $\sigma$
  counter-acts the trend of increasing [Mg/Fe] with increasing
  $\sigma$, thus the $\sigma$--[Mg/Fe] relation is substantially
  weakened when $L$ is used to parameterize galaxies instead of
  $\sigma$ }\label{cmd_mgfe}
\end{figure*}

Figure \ref{cmd_feh} shows the same slices through the color-magnitude
relation, this time with color contours indicating [Fe/H].  Again, the
trends of Figure \ref{sig_v4} are visible: [Fe/H] increases with
$\sigma$ from the lowest--$\sigma$ bins to the highest, and a range in
[Fe/H] exists within each slice in $\sigma$.  However, the variation
in [Fe/H] at fixed $\sigma$ is markedly different from the age
variation seen in Figure \ref{cmd_age}.  Lines of constant [Fe/H] run
nearly vertically in the color-magnitude diagram for fixed $\sigma$,
such that more luminous galaxies have higher [Fe/H].

The contours of [Mg/H] shown in Figure \ref{cmd_mgh} are similar to
those in [Fe/H] for the same $\sigma$ slices through the
color-magnitude relation.  In addition to the total increase in [Mg/H]
with $\sigma$, the brightest galaxies at fixed $\sigma$ have the
highest values of [Mg/H].  However, [Mg/H] shows less variation at
fixed $\sigma$ than does [Fe/H], particularly in the high--$\sigma$
bins.  This is as expected from Figure \ref{sig_v4}.  Although the
spread in [Mg/H] at fixed $\sigma$ seen here and in Figure
\ref{sig_v4} is small, Figure \ref{cmd_mgh} strongly suggests that
these variations are in fact real because they are consistent across
the $\sigma$ bins.

Finally, Figure \ref{cmd_mgfe} shows $\sigma$ slices overlaid with
contours of constant [Mg/Fe], directly comparing the differences in Fe
and Mg enrichment.  Again, [Mg/Fe] increases with increasing $\sigma$,
as expected from Figure \ref{sig_v4}.  Like [Mg/H], the variation at
fixed $\sigma$ is mild and therefore somewhat more difficult to
categorize than the variation in age and [Fe/H], but broadly [Mg/Fe]
behaves similarly to age, showing larger enhancements in fainter
galaxies at fixed $\sigma$.  Because Mg is produced by supernovae Type
II (SNe II) on short timescales while Fe is produced by supernovae
Type Ia (SNe Ia) on longer timescales, the observed overabundance of
Mg relative to the solar abundance pattern is typically interpreted to
indicate a short timescale for star formation, leaving the stellar
population of the galaxy enhanced in SN II products but not in SN Ia
products.

These maps of age, [Fe/H], [Mg/H], and [Mg/Fe] across the
color-magnitude relation confirm that quiescent galaxies form a
multi-parameter family in stellar population properties.  Age, [Fe/H],
[Mg/H], and [Mg/Fe] all increase with $\sigma$, but show variations at
fixed $\sigma$ depending on their $L$.  Of the four stellar population
parameters presented in Figures \ref{cmd_age}--\ref{cmd_mgfe}, only
age shows a strong dependence on color as well as $L$.  Variations in
age and [Mg/Fe] with $L$ at fixed $\sigma$ are similar (both increase
toward {\it fainter} galaxies) and are opposite to the variations in
[Fe/H] and [Mg/H] (both of which increase toward {\it brighter}
galaxies).  This suggests that the oldest galaxies in the universe
formed their stars over short timescales, while younger quiescent
galaxies experienced more extended star formation.

Using these stellar population maps, the differing trends with
$\sigma$ and with $L$ illustrated in Figures \ref{sig_v4} and
\ref{lum_v4} can be understood by superposing the various $\sigma$
slices of the color-magnitude relation.  Although we have seen that
age and [Fe/H] both increase with $\sigma$, Figures \ref{cmd_age} and
\ref{cmd_feh} show that age and [Fe/H] exhibit {\it opposite} behavior
from one another at fixed $\sigma$, namely that age {\it decreases}
and [Fe/H] {\it increases} for brighter galaxies at fixed $\sigma$.
Because of this difference, superposing the various $\sigma$ slices to
form the total color-magnitude relation acts to reinforce the
$\sigma$--[Fe/H] relation into an even tighter $L$--[Fe/H] relation.
At the same time, the age trends at fixed $\sigma$ counteract the
$\sigma$--age relation, resulting in almost no $L$--age relation.  The
explains why stellar population studies as a function of $L$ or
$M_*$\footnote{$M_*$ is closely related to $L$ because red sequence
galaxies span only a very limited range in $M_*/L$} find weak or
non-existant $L$--age relations (e.g.,
\citealt{kuntschner98,terlevich02,gallazzi05}), while stellar
population studies as a function of $\sigma$ turn up significant
$\sigma$--age correlations (e.g.,
\citealt{bernardi03d,thomas05,nelan05,smith07,graves07}).

In this scenario, although the $L$--[Fe/H] correlation is stronger
than the $\sigma$--[Fe/H] correlation, we interpret the
$\sigma$--[Fe/H] trend as the primary relation, with the increased
tightness of the $L$--[Fe/H] relation being caused by correlated
residuals from the $\sigma$--$L$ and $\sigma$--[Fe/H] relations.  In
the next section, we use principal components analysis to show that
the family of stellar populations in quiescent galaxies is nearly
two-dimensional, with the first dimension parameterized by $\sigma$
and the second dimension parameterized by correlated residuals from
the various $\sigma$--$L$, $\sigma$--color, and $\sigma$--stellar
population trends.

\subsection{Principal Components Analysis and Stellar Population Residuals}\label{pca}

To quantitatively explore the multi-dimensional space of global and
stellar population parameters illustrated in the previous section, we
have performed a principal components analysis (PCA; see, e.g.,
\citealt{faber73}; \citealt{trager00b}) in the seven dimensional space
parameterized by $\sigma$, $L$, color, age, [Fe/H], [Mg/H], and
[Mg/Fe].  All parameters are ``standardized'' to have a mean of zero
and a variance of one before performing the PCA.  The results are
shown in Table \ref{pca_tab}.  Although the parameter space is
nominally seven dimensional, there are only six principal components
(PCs) because [Mg/H] $\equiv$ [Mg/Fe] $+$ [Fe/H].  The first two PCs
account for 91\% of the variance in the population, indicating that
the quiescent galaxy population can be well-described by a
two-dimensional hyperplane.

The first PC (PC1) accounts for a full 70\% of the observed variation
and is comprised of positive contributions from {\it all} of the input
parameters.  This illustrates quantitatively that all the parameters
presented here are correlated with one another.  We saw in \S\ref{cmr}
that the color-magnitude relation is the result of fundamental
$\sigma$-color and $\sigma$-magnitude relations.  It therefore seems
reasonable to treat $\sigma$ as the primary global parameter behind
PC1.

\begin{deluxetable*}{lrrrrrr}[t]
\tabletypesize{\scriptsize}
\tablecaption{Principal Component Analysis\label{pca_tab}}
\tablewidth{0pt}
\tablehead{
\colhead{Parameter} &
\colhead{PC1} &
\colhead{PC2} &
\colhead{PC3} &
\colhead{PC4} &
\colhead{PC5} &
\colhead{PC6} 
}
\startdata
log $\sigma$             &$0.43$  &$ 0.03$  &$ 0.31$  &$ 0.37$  &$-0.53$ & $ 0.55$ \\
log $L$\tablenotemark{a} &$0.37$  &$-0.40$  &$ 0.04$  &$ 0.59$  &$ 0.58$ & $-0.14$ \\
$g-r$                    &$0.40$  &$ 0.23$  &$-0.62$  &$-0.25$  &$ 0.32$ & $ 0.50$ \\
log Age                  &$0.32$  &$ 0.53$  &$-0.33$  &$ 0.32$  &$-0.27$ & $-0.57$ \\
\mbox{[Fe/H]}            &$0.32$  &$-0.55$  &$-0.22$  &$-0.29$  &$-0.33$ & $-0.21$ \\
\mbox{[Mg/H]}            &$0.43$  &$-0.21$  &$ 0.14$  &$-0.40$  &$-0.08$ & $-0.21$ \\
\mbox{[Mg/Fe]}           &$0.36$  &$ 0.39$  &$ 0.59$  &$-0.34$  &$ 0.31$ & $-0.11$ \\
\hline \\
Eigenvalue               &$4.88$  &$ 1.49$  &$ 0.35$  &$ 0.18$  &$ 0.07$ & $ 0.03$ \\
Percentage of variance   &69.7    &21.3     & 5.0     & 2.6     & 1.0    &  0.4 \\
Cumulative percentage    &69.7    &91.0     &96.0     &98.6     &99.6    &100.0 \\
\enddata
\tablecomments{Parameters are ``standardized'' to have a mean of zero
  and a variance of one before performing PCA.  Although seven
  parameters are reported here, there are only six principal
  components because [Mg/H] $\equiv$ [Fe/H] $+$ [Mg/Fe].}
\tablenotetext{a}{PCA is performed using $\log L$ rather than $M_r$ to
  eliminate confusion due to the sign convention of the magnitude scale.}
\end{deluxetable*}

The second PC (PC2) encompasses another 21\% of the variance and
includes significant contributions from all parameters except for
$\sigma$.  PC2 therefore represents the correlated residuals from the
mean relations between $\sigma$ and the other parameters.  The
contributions of $g-r$, age, and [Mg/Fe] to PC2 are positive while
those of $L$, [Fe/H], and [Mg/H] are negative.  This indicates that at
fixed $\sigma$, $g-r$, age, and [Mg/Fe] are correlated with one
another and are anti-correlated with $L$, [Fe/H], and [Mg/H].  Of the
global parameters, $L$ contributes the most to PC2.  

\subsection{Stellar Population Residuals from $\sigma$ Relations}\label{residuals}

\begin{figure*}[t]
\includegraphics[width=1.0\linewidth]{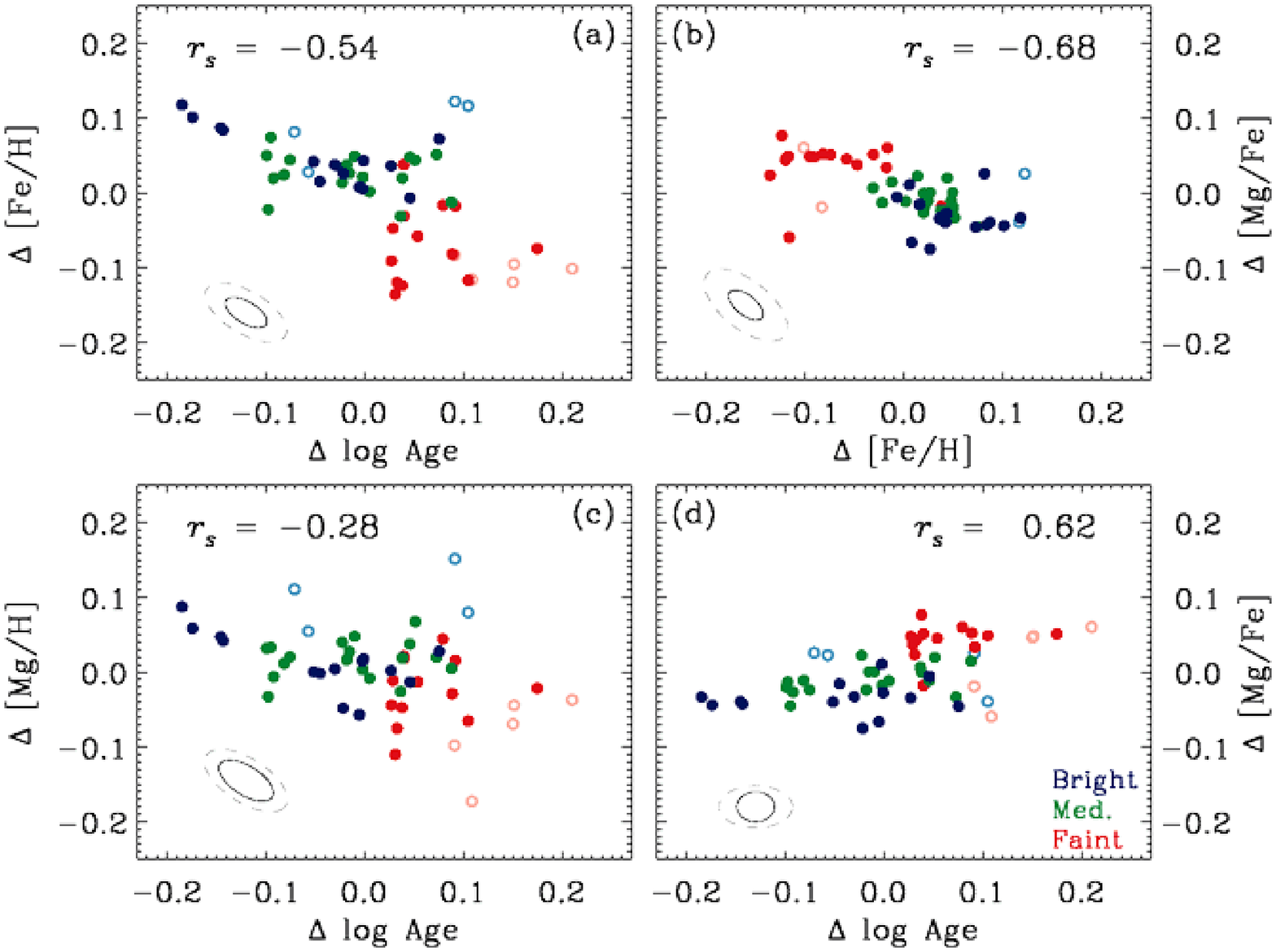}
\caption{Correlated residuals in stellar population parameters at
  fixed $\sigma$.  Residuals $\Delta \log$ Age, $\Delta$[Fe/H],
  $\Delta$[Mg/H], and $\Delta$[Mg/Fe] were computed by subtracting off
  the trends with $\sigma$ (dashed lines in Figure \ref{sig_v4}) from
  the observed values.  Data are color-coded by luminosity bin
  ($\Delta L$) as indicated.  Filled (open) circles show high (low)
  $S/N$ data, while solid (dashed) error ellipses show the
  corresponding correlated errors from stellar population modelling.
  The Spearman rank correlation coefficient ($r_s$) is indicated for
  each pair of residuals.  Residuals from the mean relations with
  $\sigma$ are related such that residuals in age and [Mg/Fe] are
  correlated with one another and are anti-correlated with residuals
  in [Fe/H] and [Mg/H].  These correlated residuals are also
  correlated with $\Delta L$, such that the brighter galaxies (blue
  points) at a given $\sigma$ have lower ages, lower [Mg/Fe], higher
  [Fe/H], and higher [Mg/H] than their fainter counterparts (red
  points).  Although the residuals are in many cases correlated in the
  same direction as the correlated errors from stellar population
  modelling, the relation between the stellar population residuals and
  $\Delta L$ could not be produced by stellar population modelling
  errors and indicates that the residual correlations are real.
}\label{resids}
\end{figure*}

Residuals in the four stellar population parameters ($\Delta \log$
age, $\Delta$[Fe/H], $\Delta$[Mg/H], and $\Delta$[Mg/Fe]) are defined
as the difference between the observed values and the mean relations
with $\sigma$, as indicated by the dashed lines in Figure
\ref{sig_v4}.  The correlated stellar population residuals associated
with PC2 are shown in Figure \ref{resids}.  In each panel, high and
low $S/N$ data are indicated by filled and open circles, respectively.
The error ellipses in the lower left corner of each panel indicate the
stastical errors from the stellar population modelling process, with
solid black (dashed gray) lines showing the median statistical errors
for high (low) $S/N$ data.  Measurement errors in the index absorption
line strengths produce correlated errors in the derived stellar
population parameters.  These are determined from Monte Carlo
simulations in \citet{graves08} and are indicated by the error ellipse
orientation.  In all panels, residuals are color-coded by luminosity
residuals; the green circles represent the central bin in $L$ at fixed
$\sigma$ (centered on the median $L$ for that $\sigma$, as in Figure
\ref{sig4_bins}) while blue and red circles indicate brighter ($\Delta
L \approx -0.8$ mag in $r$) and fainter ($\Delta L \approx +0.8$ mag
in $r$) luminosity bins.

Panel (a) shows an anti-correlation between $\Delta \log$ age and
$\Delta$[Fe/H], such that older galaxies are typically Fe-poor
compared to younger galaxies at the same $\sigma$.  This
anti-correlation at fixed $\sigma$ has been reported by numerous
previous authors (e.g.,
\citealt{worthey95,colless99,jorgensen99,trager00b}) and was
quantified by \citet{trager00b} as the ``metallicity hyperplane''.
The anti-correlation conspires to keep the color--$\sigma$ relation
tight (c.f., Figure \ref{cmd}).  In contrast to the observed
age--[Fe/H] anti-correlation at fixed $\sigma$, we saw in Figure
\ref{lum_v4} that at fixed $L$, there is significant scatter in age
but almost no scatter in [Fe/H] (although see
\citealt{poggianti01}\footnote{Unlike the results shown in Figure
\ref{lum_v4}, \citet{poggianti01} find a range of metallicity at fixed
$L$, which they observe to be anti-correlated with age.  Their results
may differ from ours because we have averaged over a large number of
galaxies in each stacked spectrum, washing out genuine [Fe/H]
variation at fixed $L$ and $\sigma$, or the observed age-metallicity
correlation in their data may be due to an underestimate of the
correlated errors in the stellar population modelling process.}).
This implies that age differences will produce color variations at
fixed $L$ which cannot be offset by correlated variations in [Fe/H].
Thus, the color-magnitude relation should show more color variation at
fixed $L$ than is observed in the color-$\sigma$ relation at fixed
$\sigma$, consistent with Figures \ref{sig_contours} and \ref{cmd}.
Furthermore, the observed color spread in the color-magnitude relation
should correlate with age, such that bluer galaxies are younger than
redder galaxies at fixed $L$.  This is consistent with Figure
\ref{cmd_age} and also with the results of \citet{cool06}.

In addition to the age--[Fe/H] anti-correlation evident in Figure
\ref{resids}a, a similar anti-correlation is present between age and 
[Mg/H] (Figure \ref{resids}c), although it is significantly
weaker.  A further anti-correlation is observed between [Fe/H] and
[Mg/Fe] (Figure \ref{resids}b), such that Fe-poor galaxies are
more Mg-enhanced than their comparitively Fe-rich counterparts at
fixed $\sigma$.  Finally, Figure \ref{resids}d shows a positive
correlation between age and [Mg/Fe], such that older galaxies are
more Mg-enhanced with respect to the solar abundance pattern than
their younger counterparts at fixed $\sigma$.  

In panels a--c, the slopes of the observed anti-correlations are
similar to those expected from correlated errors in stellar population
modelling, as indicated by the error ellipses.  However, the observed
anti-correlations cannot be due to correlated statistical errors for
several reasons.  Firstly, the typical statistical errors are too
small to produce the observed spread in stellar population properties.
More conclusively, if the observed anti-correlations were due to
measurement errors, they would not be systematically correlated with
$\Delta L$.  The observed $L$--dependence of stellar population
properties at fixed $\sigma$ argues strongly that the anti-correlation
is real.  In appendix \ref{young_pops}, we verify that the observed
anti-correlations are not due to {\it systematic} effects in the
stellar population modelling process caused by using single burst
models to fit integrated galaxy spectra, which are almost certainly not
single burst populations.

It is evident from Figure \ref{resids} that, not only are the various
stellar population residuals correlated with one another, they are
also strongly correlated with $\Delta L$, as quantified by the PCA
analysis in the previous section.  Collecting all parameters together,
the residual trends can be summarized as follows: at fixed $\sigma$,
galaxies fainter than the median are older, Fe-poor, somewhat Mg-poor,
and more Mg-enhanced than typical galaxies at that $\sigma$, while
galaxies brighter than the median are younger, Fe-rich, somewhat
Mg-rich, and less Mg-enhanced.

\section{Discussion}\label{discussion}

We have shown that (1) stellar population age, [Fe/H], [Mg/H], and
[Mg/Fe] all increase with increasing galaxy $\sigma$ (Figure
\ref{sig_v4}), and (2) that the star formation histories of galaxies
vary systematically at fixed $\sigma$ such that fainter galaxies are
older, Fe-poor, and Mg-enhanced compared to their bright counterparts
at the same $\sigma$ (Figure \ref{resids}).  The first of these
results is in agreement with a substantial number of earlier works
(e.g, \citealt{bernardi03d}; \citealt{thomas05}; \citealt{nelan05};
\citealt{smith07}; \citealt{graves07}).  In particular, the increase
in age with $\sigma$ is consistent with ``archeological downsizing''
\citep{thomas05}, in which massive galaxies form their stars
predominantly at early times, while lower mass galaxies show evidence
of younger stellar populations.

The anti-correlation of age and [Fe/H] at fixed $\sigma$ shown in this
analysis is not a new result and is in agreement with the metallicity
hyperplane of \citet{trager00b} and with the recent work of
\citet{smith07-iaus245} for individual galaxies.  However, an
important new result of this work is that {\it this anti-correlation
is correlated with galaxy luminosity}, such that the more luminous
galaxies at fixed $\sigma$ are the younger, Fe-rich galaxies while the
fainter galaxies are older and more Fe-poor.  This confirms that the
observed anti-correlation of age and [Fe/H] must be genuine, rather
than an effect of correlated stellar population modelling errors.  

The co-variation of age, [Fe/H], and [Mg/Fe] we observe at fixed
$\sigma$ cannot be driven by large-scale environment.  The age--[Fe/H]
anti-correlation is observed separately in the \citet{trager00b} data,
which include only field galaxies, and in the \citet{smith07-iaus245}
data, which include only cluster galaxies.  It is also observed in our
stacked spectra, which average over all environments.  

The observed co-variation of age, [Fe/H], and [Mg/Fe] is qualitatively
consistent with a scenario in which all galaxies at the same $\sigma$
start forming stars at the same $z$ but the duration of star formation
varies between galaxies.  At fixed $\sigma$, galaxies with short star
formation timescales end star formation early and consequently have
older mean stellar ages.  They also have less time for enrichment in
SN Ia products, resulting in lower [Fe/H] and higher [Mg/Fe].  To the
extent that $\sigma$ measures the depth of the gravitational
potential, the loss of metals through winds should be consistent for
all galaxies at the same $\sigma$ if they experience comparable
galactic winds. Thus early truncation of star formation should have
little effect on [Mg/H] but should result in a genuine underabundance
of [Fe/H], as observed.  To also be consistent with the general trends
of Figure \ref{sig_v4}, these variations in the duration of star
formation must be superimposed on a staged star formation model,
similar to the \citet{noeske07b} model, where the redshift at which
star formation peaks and the duration of star formation depend on
$\sigma$.  Unlike the Noeske model, the data presented here
additionally require a spread in star formation duration at fixed
$\sigma$.  A quantitative test of such a model requires the
construction of chemical evolution models, which we defer to future
work.  Furthermore, this scenario postulates a range of star formation
timescales at fixed $\sigma$ but does not propose a physical mechanism
to produce the observed variation.

A key result of this analysis is that the co-variation of age, [Fe/H],
and [Mg/Fe] at fixed $\sigma$ also correlates with galaxy luminosity,
such that the brighter galaxies are younger, Fe-rich, and less
$\alpha$-enhanced than the fainter galaxies at the same $\sigma$.  If
the above scenario is correct, the more luminous galaxies at fixed
$\sigma$ would be those which experienced more extended star formation
histories.  This at first appears to be a trivial result, since
galaxies with more extended star formation and younger mean ages are
expected to have lower $M/L$ and therefore higher $L$ at fixed
$\sigma$.  However, the differences in $M/L$ predicted from stellar
population models are too small to account for the observed variation
in $L$.  This will be shown in detail in future papers in this series,
but can also be inferred from Figure \ref{sig_contours}.  This figure
shows that the range in galaxy colors at fixed $\sigma$ is small,
suggesting that the range of $M/L$ is likewise small.  In fact, the
variation in $L$ at fixed $\sigma$ turns out to be dominated by
variation in total stellar mass at fixed $\sigma$, while $M/L$
variations contribute only a modest amount.  Thus any physical
mechanism invoked to explain the observed co-variation in stellar
population properties at fixed $\sigma$ must also reproduce differing
total stellar mass content in the galaxies.

\section{Conclusions}\label{conclusions}

This analysis has used very high $S/N$ stacked spectra of $\sim16,000$
SDSS early type galaxies to map out variations in stellar population
age, [Fe/H], [Mg/H], and [Mg/Fe] with galaxy $\sigma$, $L$, and color.
In addition to studying the mean trends in stellar populations with
each of these three global properties individually, we have mapped out
variations in age, [Fe/H], [Mg/H], and [Mg/Fe] in the three
dimensional $\sigma$--$L$--color parameter space.  This allows us to
understand the differing behavior of stellar populations as a function
of $\sigma$ versus $L$, and to explore the systematic variations in
stellar populations at fixed $\sigma$.

We find the following results for quiescent galaxies:
\begin{list}{}{}
\item[1.] Luminosity and $\sigma$ both increase with galaxy mass, yet
  these ``size'' measures are not the same.  Fundamental stellar
  population variables such as age, [Fe/H], [Mg/H], and [Mg/Fe] scale
  differently with respect to $L$ and to $\sigma$, and also with
  respect to galaxy color.
\item[2.] Higher $\sigma$ galaxies are typically more luminous and
  redder than lower $\sigma$ galaxies.  However, at fixed $\sigma$,
  there is no color--magnitude relation, despite the substantial
  variation in $M_r$.  The color-magnitude relation of passive
  galaxies is therefore a result of combining the $\sigma$--$L$ and
  $\sigma$--color relations, in agreement with \citet{bernardi05}.
\item[3.] Mean luminosity-weighted stellar population age, [Fe/H],
  [Mg/H], and [Mg/Fe] all correlate with $\sigma$ such that galaxies
  with higher $\sigma$ tend to be older and have higher [Fe/H],
  [Mg/H], and [Mg/Fe].  At fixed $\sigma$, the spread in [Mg/H] and
  [Mg/Fe] is small, while both [Fe/H] and age show significant spread.
\item[4.] At fixed $\sigma$, brighter galaxies have lower age, higher
  [Fe/H], slightly higher [Mg/H], and lower [Mg/Fe] than their fainter
  counterparts.  The anti-correlation of age and [Fe/H] conspires to
  keep the color--magnitude relation flat at fixed $\sigma$ and
  contributes to the overall narrowness of the color--$\sigma$
  relation.
\item[5.] At fixed $\sigma$, age is also correlated with color such
  that younger galaxies are both brighter and bluer at fixed $\sigma$
  than the fainter, redder, older galaxies.  In contrast, [Fe/H]
  and [Mg/H] do not vary systematically with color, while [Mg/Fe]
  varies only mildly with color.
\item[6.] The variation in stellar population properties with $L$ at
  fixed $\sigma$ acts to reinforce the $\sigma$--[Fe/H] trend,
  resulting in a strong and tight $L$--[Fe/H] correlation.  In
  contrast, the $\sigma$--age and $\sigma$--[Mg/Fe] correlations are
  opposed by the variations with $L$ at fixed $\sigma$, leaving only
  weak $L$--age and $L$--[Mg/Fe] correlations.  In a similar way,
  residuals in color correlate with stellar population properties such
  that there is a strong color--age trend but only weak color--[Fe/H]
  and color--[Mg/Fe] trends.  Age correlates most closely with color,
  [Fe/H] correlates most closely with $L$, and [Mg/H] and [Mg/Fe]
  correlate most closely with $\sigma$.  Only $\sigma$ (and not $L$ or
  color) correlates strongly with all four stellar population
  properties.
\item[7.] Trends in age, [Fe/H], [Mg/H], and [Mg/Fe] at fixed $\sigma$
  are likely not driven by environment.  The age--[Fe/H]
  anti-correlation at fixed $\sigma$ is detected in samples of
  individual galaxies which reside in similar environments (e.g., the
  field galaxies in \citealt{trager00b} and the cluster galaxies in
  \citealt{smith07-iaus245}) and also in our sample of stacked
  spectra, which average over all environments.  
\end{list}

The variations in stellar population properties at fixed $\sigma$
presented here illustrate that the narrow, seemingly one-dimensional
color magnitude relation of quiescent galaxies conceals an underlying
set of star formation histories that populate a two-parameter family.
Not only do the star formation histories of galaxies vary
systematically with their $\sigma$, but there is clearly a range of
processes at work in galaxies of the same $\sigma$.  These result in
an anti-correlation between galaxy age and [Fe/H] at fixed $\sigma$
and a weaker but positive age--[Mg/Fe] correlation.  Moreover, these
properties are linked to the present day luminosities of the galaxies,
such that brighter galaxies are younger, more metal-rich, and less
$\alpha$-enhanced than fainter galaxies at the same $\sigma$.  The
companion papers in this series will explore in more detail the
connection between galaxy structure and galaxy stellar populations,
providing evidence for the ways in which the star formation history of
a galaxy is linked to its mass assembly and structural evolution.

\acknowledgements

The authors would like to thank Renbin Yan for providing the emission
line measurements used to identify the sample of early-type galaxies
used here, as well as an anonymous referee for provided valuable
feedback.  This work was supported by National Science Foundation
grant AST 05-07483.

Funding for the creation and distribution of the SDSS Archive has been
provided by the Alred P. Sloan Foundation, the Participating
Institutions, the National Aeronautics and Space Administration, the
National Science Foundation, the US Department of Energy, the Japanese
Monbukagakusho, and the Max-Planck Society. The SDSS Web site is
http://www.sdss.org/.

The SDSS is managed by the Astrophysical Research Consortium (ARC) for
the Participating Institutions. The Participating Institutions are the
University of Chicago, Fermilab, the Institute for Advanced Study, the
Japan Participation Group, the Johns Hopkins University, the Korean
Scientist Group, Los Alamos National Laboratory, the
Max-Planck-Institute for Astronomy (MPIA), the Max-Planck-Institute
for Astrophysics (MPA), New Mexico State University, University of
Pittsburgh, University of Portsmouth, Princeton University, the United
States Naval Observatory, and the University of Washington.

\bibliographystyle{apj}
\bibliography{apj-jour,myrefs}

\appendix

\section{Bulge $+$ Disk vs. de Vaucouleurs Magnitudes and Colors}
\label{dev_colors}

This analysis is based on magnitudes and colors measured from de
Vaucouleur fits to the galaxy light profiles.  However, given that a
fraction of the galaxies in the sample have a visible disk component it
is important to verify that the de Vaucouleurs magnitudes and colors
are a reasonable approximation to the total galaxy magnitudes and
colors.  

The Petrosian magnitudes measured by the SDSS pipeline have larger
errors (typically 5\% in the $r$-band) than the de Vaucouleurs model
fits (typically 2\% in the $r$-band) as determined by repeat
observations of targeted galaxies.  However, they have the benefit of
being independent of the galaxy light profile.  We compare Petrosian
$r$-band magnitudes to the de Vaucouleurs magnitudes used in our
analysis and find that the de Vaucouleurs magnitudes are
systematically brighter by $<0.1$ mag (consistent with
\citealt{blanton01}), with rms scatter of $0.04$ mag.  This scatter is
much smaller than the 0.8 mag bin widths used to sort and stack galaxy
spectra and should therefore have little impact on the analysis
presented here.

\begin{figure}
\includegraphics[width=1.0\linewidth]{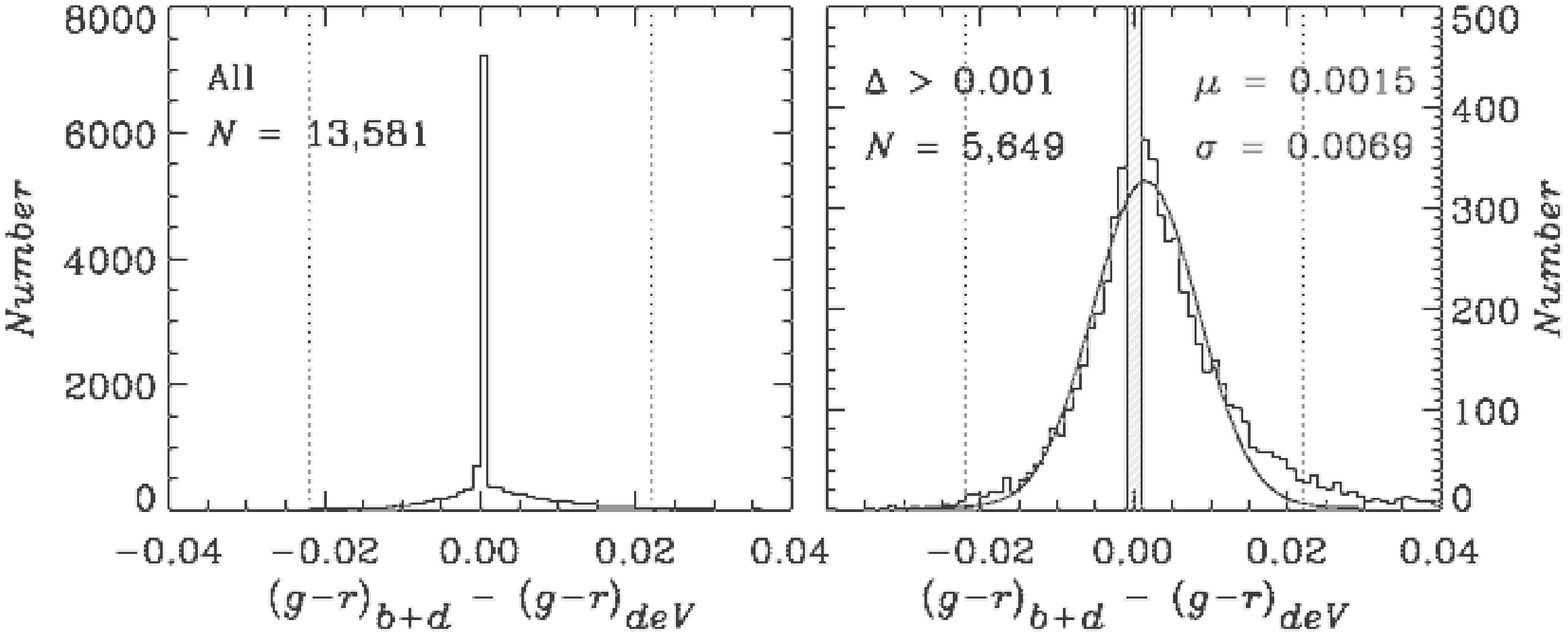}
\caption{Distribution of $g-r$ color differences between composite
  bulge $+$ disk and de Vaucouleurs model light profiles for our
  sample galaxies.  In the bottom panel, the galaxies with color
  differences $< 0.001$ mag have been removed, showing the nearly
  Gaussian distribution of the remaining color differences.  Overall,
  $>97$\% of the color differences are less than the $0.022$ mag rms
  error in the $g-r$ colors measured from SDSS photometry (dotted
  lines).  We conclude that using magnitudes and colors derived from
  de Vaucouleurs fits to the galaxy light profiles does not
  substantially affect the results presented here.
}\label{color_hists}
\end{figure}

Color effects may be more important, both because the color
differences observed in our sample galaxies are intrinsically small
and because disk galaxies may exhibit substantial color gradients.  We
combine the separate de Vaucouleurs and exponential fits to galaxy
light profiles, weighting each component by the fractional
contribution of each component (the SDSS pipeline value {\it fracDeV})
to produce composite bulge $+$ disk galaxy magnitudes and colors.  We
compare the $g-r$ colors from the bulge $+$ disk models to the de
Vaucouleurs profile colors in Figure \ref{color_hists}.  In the top
panel we show the distribution of color differences for all galaxies
that go into our stacked spectra (i.e., after the color- and
magnitude-outliers have been removed from each $\sigma$ bin).  The
majority of galaxies show $< 0.001$ mag difference between the
composite and de Vaucouleur color values because the galaxies are
highly bulge-dominated.  Removing these and examining the distribution
of color differences for galaxies with color differences {\it larger}
than 0.001 mag (bottom panel), we find that the distribution is
roughly gaussian, with $\mu = 0.0015$ mag and $\sigma = 0.0069$ mag.

Given that the rms error for $g-r$ color photometry in SDSS is $\sim
2$\% (0.022 mag, dotted lines in Figure \ref{color_hists}), the color
differences from the different model light profiles are within this
error for $>97$\% of our galaxy sample.  We therefore conclude that
the use of de Vaucouleurs model fits for the magnitudes and colors in
this analysis does not have a significant effect on the results.

\section{The Effect of Young Sub-Populations}\label{young_pops}

Because the observed age--[Fe/H] anti-correlation at fixed $\sigma$ is
in the same sense as the age-metallicity degeneracies that plague
stellar population studies, it is necessary to confirm that {\it
  systematic} errors in the modelling process are not producing a
spurious anti-correlation.  We showed in \S\ref{resids} that the
age--[Fe/H] anti-correlation at fixed $\sigma$ is not due to random
observational errors.  However, the models used in this work are
single stellar population (SSP) models, whereas galaxies are composite
populations with extended star formation histories.  The mean age
measured from an SSP model is a mean luminosity-weighted age for all
the stars contributing to the total light.  Because young populations
have lower mass-to-light ratios and stronger Balmer absorption lines,
a sub-population of intermediate age stars (either a younger
sub-population within individual galaxies or a subset of significantly
younger galaxies in the bin) will contribute disproportionately to the
total galaxy light, skewing the SSP age to low values.  If the young
sub-population greatly strengthens the Balmer absorption lines without
changing the Fe absorption lines much, the resulting model grid
inversion may infer higher metallicities just due to the stronger
Balmer lines rather than a genuine difference in metallicity.

\begin{figure}
\includegraphics[width=0.5\linewidth]{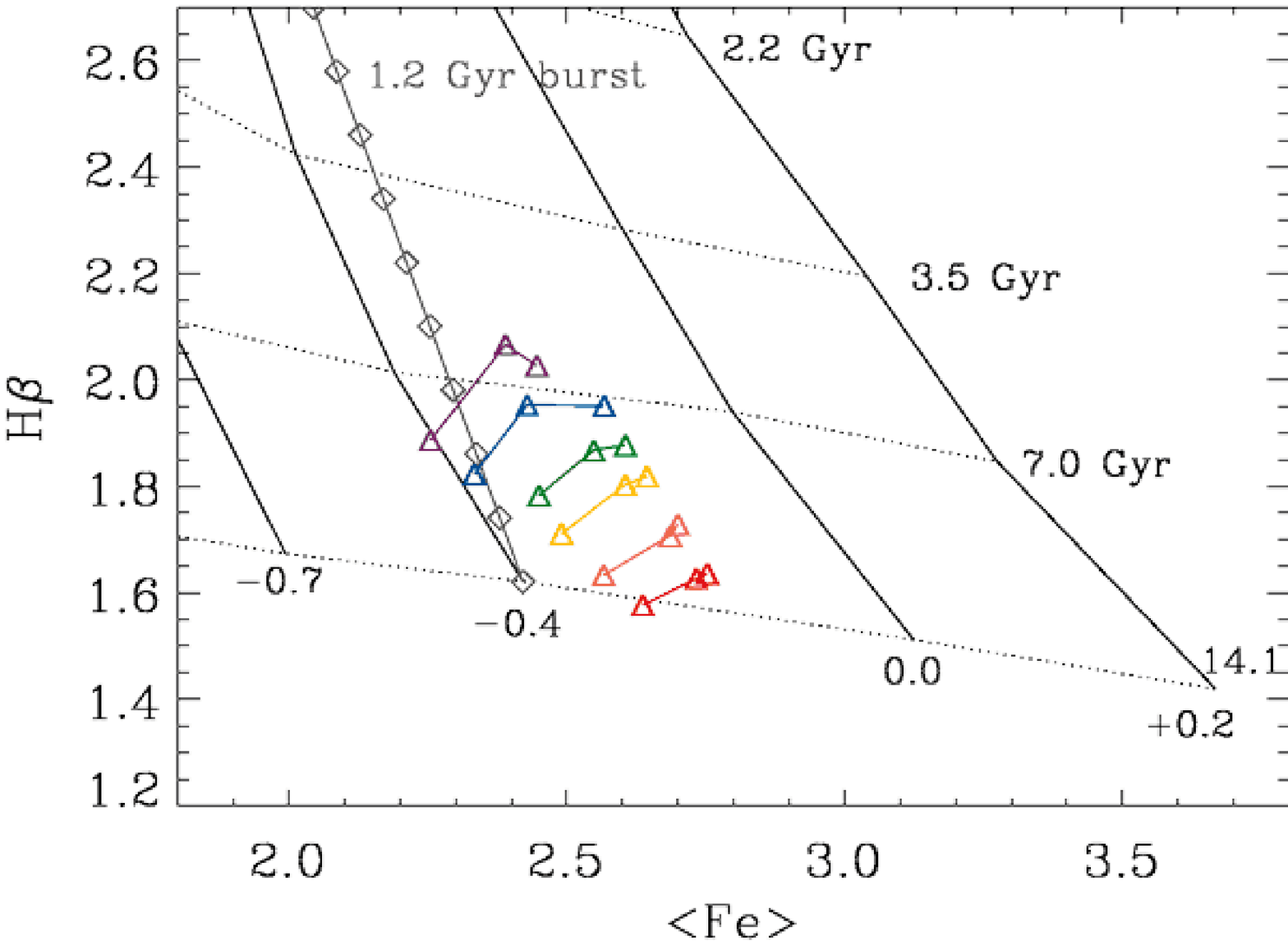}
\caption{Stellar population model grid showing the effect of adding a
  young burst to an underlying old population.  The model grid is for
  a solar-scale abundance pattern model from \citet{schiavon07}.
  Solid (dotted) lines connect models with the same [Fe/H] (age).  The
  gray track shows the effect of taking an old population with age =
  14.1 Gyr and [Fe/H] = -0.4, and adding to it varying fractions of a
  1.2 Gyr burst with the same [Fe/H].  Gray diamonds indicate the
  fraction of the total luminosity at 5000{\AA} provided by the young
  burst, starting at 0\% at the bottom and increasing in increments of
  5\%.  Index values from the stacked spectra are overplotted as
  triangles, with colors indicating the $\sigma$ range of the stacked
  spectrum, as in Figure \ref{sig_contours} (purple shows the lowest
  $\sigma$ bin, red the highest $\sigma$ bin).  The three data points
  at fixed $\sigma$ show the three different bins in $L$.  The
  age--[Fe/H] anticorrelation observed at fixed $\sigma$ cannot simply
  be due to the presence of a young sub-population in the brighter
  galaxies because the trajectory of a young burst at constant [Fe/H]
  is nearly orthogonal to the observed co-variation of
  $\langle$Fe$\rangle$ and H$\beta$ at fixed $\sigma$.  The brighter
  galaxies must have higher average [Fe/H] than their faint
  counterparts at the same $\sigma$, in addition to having younger
  mean ages.  }\label{grid}
\end{figure}

This is in the correct sense to account for the anti-correlation we
see between galaxy age and [Fe/H] at fixed $\sigma$.  However, it
alone cannot account for metallicity differences at fixed $\sigma$, as
illustrated in Figure \ref{grid}.  This figure shows a stellar
population model grid for the $\langle$Fe$\rangle$ and H$\beta$
indices.  Solid lines connect models of constant [Fe/H], while dotted
lines connect models of constant age.  Triangles show the index
measurements for stacked spectra in this work, with colors indicating
the $\sigma$ range represented (purple shows the lowest $\sigma$ bin,
red the highest $\sigma$ bin, as in Figure \ref{sig_contours}).  We
have averaged over the three different color bins so the three points
shown for each $\sigma$ bin represent three different sub-bins in $L$.
The gray track with diamonds shows the effect of adding a young
sub-population to an underlying old population.  The old population
has age = 14.1 Gyr and [Fe/H] = -0.4, while the young sub-population
has the same [Fe/H] and age = 1.2 Gyr.  Gray diamonds indicate the
fraction of the total luminosity at 5000 {\AA} provided by the burst,
starting at 0\% at the bottom and increasing in increments of 5\%.

It is clear that the trajectory in index-index space covered by
composite burst models with fixed [Fe/H] is very different from the
observed co-variation in galaxy age and [Fe/H].  It is true that the
[Fe/H] inferred for a composite burst would be skewed slightly high,
since the gray track in Figure \ref{grid} lies consistently to the
right of the [Fe/H] = -0.4 grid line despite the fact that both
sub-populations in the composite models have [Fe/H] = -0.4.  However,
the effect is small compared to the observed variation in
$\langle$Fe$\rangle$ at fixed $\sigma$ in the galaxy sample.  The data
therefore require genuine variations in mean [Fe/H] at fixed $\sigma$,
in addition to age variations.  

\section{}

\begin{deluxetable}{ccccccccc}
\tabletypesize{\scriptsize}
\tablecaption{Properties of Galaxy Bins and Stacked Spectra\label{bin_tab}}
\tablewidth{0pt}
\tablehead{
\colhead{} &
\colhead{} &
\colhead{} &
\colhead{} &
\colhead{Median} &
\colhead{Median} &
\colhead{Median} &
\colhead{} \\
\colhead{} &
\colhead{$\sigma$ bin} &
\colhead{$M_r$ bin} &
\colhead{$g-r$ bin} &
\colhead{$\log \sigma$} &
\colhead{$M_r$} &
\colhead{$(g-r)_{deV}$\tablenotemark{*}} &
\colhead{Number\tablenotemark{**}} &
\colhead{$S/N$\tablenotemark{\dag}} \\
\colhead{} &
\colhead{} &
\colhead{} &
\colhead{} &
\colhead{(km s$^{-1}$)} &
\colhead{(mag)} &
\colhead{(mag)} &
\colhead{} &
\colhead{({\AA}$^{-1}$)} 
 }
\startdata
 1&$1.86 < \log \sigma < 2.00$&$-21.29 < M_r < -20.49$&$ 0.665$--$ 0.697$&1.95   &-20.76    &0.685 &  52  &149    \\     
 2&                           &                       &$ 0.697$--$ 0.729$&1.97   &-20.64    &0.712 &  85  &170    \\     
 3&                           &                       &$ 0.729$--$ 0.761$&1.97   &-20.65    &0.741 &  57  &128    \\     
 4&                           &$-20.49 < M_r < -19.69$&$ 0.665$--$ 0.697$&1.94   &-20.05    &0.684 & 177  &219    \\     
 5&                           &                       &$ 0.697$--$ 0.729$&1.95   &-20.08    &0.713 & 390  &325    \\     
 6&                           &                       &$ 0.729$--$ 0.761$&1.95   &-20.12    &0.741 & 233  &252    \\     
 7&                           &$-19.69 < M_r < -18.89$&$ 0.665$--$ 0.697$&1.92   &-19.40    &0.685 &  55  &113    \\     
 8&                           &                       &$ 0.697$--$ 0.729$&1.94   &-19.46    &0.715 & 122  &171    \\     
 9&                           &                       &$ 0.729$--$ 0.761$&1.95   &-19.46    &0.740 &  63  &130    \\     
10&$2.00 < \log \sigma < 2.09$&$-21.54 < M_r < -20.74$&$ 0.680$--$ 0.712$&2.05   &-20.94    &0.699 &  81  &189    \\     
11&                           &                       &$ 0.712$--$ 0.744$&2.07   &-20.92    &0.728 & 176  &288    \\     
12&                           &                       &$ 0.744$--$ 0.776$&2.07   &-20.95    &0.753 &  87  &188    \\     
13&                           &$-20.74 < M_r < -19.94$&$ 0.680$--$ 0.712$&2.05   &-20.29    &0.702 & 290  &317    \\     
14&                           &                       &$ 0.712$--$ 0.744$&2.05   &-20.32    &0.727 & 724  &511    \\     
15&                           &                       &$ 0.744$--$ 0.776$&2.06   &-20.33    &0.755 & 357  &360    \\     
16&                           &$-19.94 < M_r < -19.14$&$ 0.680$--$ 0.712$&2.04   &-19.75    &0.702 &  93  &158    \\     
17&                           &                       &$ 0.712$--$ 0.744$&2.05   &-19.78    &0.728 & 166  &235    \\     
18&                           &                       &$ 0.744$--$ 0.776$&2.05   &-19.76    &0.753 &  80  &164    \\     
19&$2.09 < \log \sigma < 2.18$&$-21.84 < M_r < -21.04$&$ 0.693$--$ 0.725$&2.15   &-21.31    &0.713 & 170  &355    \\     
20&                           &                       &$ 0.725$--$ 0.757$&2.15   &-21.24    &0.741 & 376  &488    \\     
21&                           &                       &$ 0.757$--$ 0.789$&2.15   &-21.23    &0.768 & 142  &287    \\     
22&                           &$-21.04 < M_r < -20.24$&$ 0.693$--$ 0.725$&2.13   &-20.63    &0.715 & 483  &479    \\     
23&                           &                       &$ 0.725$--$ 0.757$&2.14   &-20.63    &0.741 &1138  &731    \\     
24&                           &                       &$ 0.757$--$ 0.789$&2.14   &-20.61    &0.767 & 552  &507    \\     
25&                           &$-20.24 < M_r < -19.44$&$ 0.693$--$ 0.725$&2.12   &-20.07    &0.715 & 174  &231    \\     
26&                           &                       &$ 0.725$--$ 0.757$&2.13   &-20.05    &0.740 & 376  &367    \\     
27&                           &                       &$ 0.757$--$ 0.789$&2.13   &-20.08    &0.767 & 141  &235    \\     
28&$2.18 < \log \sigma < 2.27$&$-22.23 < M_r < -21.43$&$ 0.705$--$ 0.737$&2.23   &-21.63    &0.726 & 206  &427    \\     
29&                           &                       &$ 0.737$--$ 0.769$&2.24   &-21.62    &0.753 & 398  &568    \\     
30&                           &                       &$ 0.769$--$ 0.801$&2.24   &-21.64    &0.779 & 150  &337    \\     
31&                           &$-21.43 < M_r < -20.63$&$ 0.705$--$ 0.737$&2.21   &-21.05    &0.727 & 420  &520    \\     
32&                           &                       &$ 0.737$--$ 0.769$&2.22   &-21.03    &0.753 &1057  &844    \\     
33&                           &                       &$ 0.769$--$ 0.801$&2.23   &-21.02    &0.780 & 533  &596    \\     
34&                           &$-20.63 < M_r < -19.83$&$ 0.705$--$ 0.737$&2.21   &-20.36    &0.728 & 221  &301    \\     
35&                           &                       &$ 0.737$--$ 0.769$&2.22   &-20.38    &0.754 & 482  &485    \\     
36&                           &                       &$ 0.769$--$ 0.801$&2.22   &-20.44    &0.779 & 166  &283    \\     
37&$2.27 < \log \sigma < 2.36$&$-22.66 < M_r < -21.86$&$ 0.718$--$ 0.750$&2.31   &-22.11    &0.741 & 138  &385    \\     
38&                           &                       &$ 0.750$--$ 0.782$&2.32   &-22.08    &0.766 & 299  &536    \\     
39&                           &                       &$ 0.782$--$ 0.814$&2.33   &-22.05    &0.790 & 111  &334    \\     
40&                           &$-21.86 < M_r < -21.06$&$ 0.718$--$ 0.750$&2.29   &-21.51    &0.742 & 250  &476    \\     
41&                           &                       &$ 0.750$--$ 0.782$&2.31   &-21.48    &0.766 & 671  &785    \\     
42&                           &                       &$ 0.782$--$ 0.814$&2.32   &-21.44    &0.790 & 246  &462    \\     
43&                           &$-21.06 < M_r < -20.26$&$ 0.718$--$ 0.750$&2.29   &-20.72    &0.743 & 113  &258    \\     
44&                           &                       &$ 0.750$--$ 0.782$&2.30   &-20.79    &0.766 & 313  &443    \\     
45&                           &                       &$ 0.782$--$ 0.814$&2.30   &-20.82    &0.791 & 149  &325    \\     
46&$2.36 < \log \sigma < 2.50$&$-23.02 < M_r < -22.22$&$ 0.732$--$ 0.764$&2.39   &-22.37    &0.757 &  45  &242    \\     
47&                           &                       &$ 0.764$--$ 0.796$&2.40   &-22.39    &0.780 & 116  &380    \\     
48&                           &                       &$ 0.796$--$ 0.828$&2.40   &-22.46    &0.803 &  52  &257    \\     
49&                           &$-22.22 < M_r < -21.42$&$ 0.732$--$ 0.764$&2.38   &-21.76    &0.756 &  88  &319    \\     
50&                           &                       &$ 0.764$--$ 0.796$&2.38   &-21.86    &0.781 & 228  &497    \\     
51&                           &                       &$ 0.796$--$ 0.828$&2.40   &-21.96    &0.803 &  88  &320    \\     
52&                           &$-21.42 < M_r < -20.62$&$ 0.732$--$ 0.764$&2.38   &-21.11    &0.755 &  54  &218    \\     
53&                           &                       &$ 0.764$--$ 0.796$&2.38   &-21.11    &0.779 & 109  &304    \\     
54&                           &                       &$ 0.796$--$ 0.828$&2.40   &-21.06    &0.810 &  38  &187    \\     
\enddata
\tablenotetext{*}{$g-r$ color as measured from the total de
  Vaucouleurs galaxy photometry}
\tablenotetext{**}{Total number of galaxies in the stacked spectrum.}
\tablenotetext{\dag}{Effective median $S/N$ of the stacked spectrum.}
\end{deluxetable}

\begin{deluxetable}{cc@{ $\pm$ }lc@{ $\pm$ }lc@{ $\pm$ }lr@{ $\pm$ }lr@{ $\pm$ }lr@{ $\pm$ }lr@{ $\pm$ }l}
\tabletypesize{\scriptsize}
\tablecaption{Lick Indices and Stellar Population Properties of
  Stacked Spectra\label{index_tab}}
\tablewidth{0pt}
\tablehead{
\colhead{} &
\multicolumn{2}{c}{H$\beta$} &
\multicolumn{2}{c}{$\langle$Fe$\rangle$} &
\multicolumn{2}{c}{Mg $b$} &
\multicolumn{2}{c}{Age} &
\multicolumn{2}{c}{[Fe/H]} &
\multicolumn{2}{c}{[Mg/H]} &
\multicolumn{2}{c}{[Mg/Fe]} \\
\colhead{} &
\multicolumn{2}{c}{({\AA})} &
\multicolumn{2}{c}{({\AA})} &
\multicolumn{2}{c}{({\AA})} &
\multicolumn{2}{c}{(Gyr)} &
\multicolumn{2}{c}{(dex)} &
\multicolumn{2}{c}{(dex)} &
\multicolumn{2}{c}{(dex)} 
}
\startdata
 1  &2.10 &0.05   &2.34 &0.06   &3.16 &0.05  & 5.5 &0.5  &-0.22 &0.03  &-0.08 &0.04  &0.14 &0.03\\
 2  &2.09 &0.04   &2.42 &0.05   &3.38 &0.04  & 5.5 &0.4  &-0.16 &0.03  &-0.01 &0.03  &0.15 &0.02\\
 3  &1.88 &0.06   &2.58 &0.07   &3.78 &0.06  & 8.0 &0.8  &-0.12 &0.03  & 0.03 &0.05  &0.15 &0.03\\
 4  &2.17 &0.03   &2.39 &0.04   &2.99 &0.03  & 5.0 &0.3  &-0.18 &0.02  &-0.11 &0.03  &0.07 &0.02\\
 5  &2.07 &0.03   &2.35 &0.03   &3.20 &0.03  & 6.0 &0.4  &-0.24 &0.02  &-0.10 &0.03  &0.14 &0.02\\
 6  &1.96 &0.03   &2.43 &0.04   &3.41 &0.03  & 7.1 &0.4  &-0.21 &0.02  &-0.07 &0.03  &0.14 &0.02\\
 7  &1.98 &0.06   &2.24 &0.07   &2.88 &0.06  & 7.5 &0.9  &-0.34 &0.04  &-0.25 &0.05  &0.09 &0.04\\
 8  &1.89 &0.04   &2.23 &0.05   &3.14 &0.04  & 8.7 &0.7  &-0.38 &0.02  &-0.22 &0.03  &0.16 &0.02\\
 9  &1.80 &0.06   &2.30 &0.06   &3.41 &0.05  &10.3 &1.1  &-0.35 &0.03  &-0.17 &0.05  &0.18 &0.05\\
10  &2.16 &0.04   &2.49 &0.04   &3.34 &0.04  & 4.8 &0.3  &-0.10 &0.02  & 0.02 &0.03  &0.12 &0.02\\
11  &1.92 &0.03   &2.51 &0.03   &3.48 &0.03  & 7.5 &0.4  &-0.17 &0.02  &-0.04 &0.03  &0.13 &0.02\\
12  &1.77 &0.04   &2.70 &0.04   &3.92 &0.04  & 9.6 &0.7  &-0.09 &0.04  & 0.03 &0.05  &0.12 &0.02\\
13  &2.07 &0.03   &2.35 &0.03   &3.18 &0.03  & 5.9 &0.4  &-0.24 &0.02  &-0.10 &0.03  &0.14 &0.02\\
14  &1.96 &0.02   &2.47 &0.03   &3.44 &0.03  & 7.1 &0.3  &-0.18 &0.01  &-0.05 &0.02  &0.13 &0.01\\
15  &1.82 &0.03   &2.47 &0.03   &3.61 &0.03  & 9.1 &0.5  &-0.22 &0.02  &-0.05 &0.03  &0.17 &0.02\\
16  &1.85 &0.05   &2.33 &0.05   &3.10 &0.04  & 9.3 &0.8  &-0.33 &0.03  &-0.24 &0.04  &0.09 &0.03\\
17  &1.85 &0.03   &2.30 &0.04   &3.42 &0.03  & 9.3 &0.6  &-0.33 &0.02  &-0.13 &0.03  &0.20 &0.02\\
18  &1.77 &0.05   &2.37 &0.05   &3.56 &0.05  &10.3 &0.9  &-0.31 &0.03  &-0.11 &0.04  &0.20 &0.03\\
19  &2.05 &0.02   &2.58 &0.02   &3.60 &0.02  & 5.6 &0.3  &-0.08 &0.02  & 0.06 &0.02  &0.14 &0.01\\
20  &1.85 &0.02   &2.54 &0.02   &3.50 &0.02  & 8.4 &0.3  &-0.17 &0.01  &-0.05 &0.02  &0.12 &0.01\\
21  &1.73 &0.03   &2.70 &0.03   &3.97 &0.03  &10.1 &0.5  &-0.11 &0.01  & 0.03 &0.03  &0.14 &0.02\\
22  &2.00 &0.02   &2.49 &0.02   &3.59 &0.02  & 6.6 &0.3  &-0.14 &0.01  & 0.02 &0.02  &0.16 &0.02\\
23  &1.86 &0.02   &2.50 &0.02   &3.67 &0.02  & 8.4 &0.3  &-0.18 &0.01  &-0.01 &0.01  &0.17 &0.01\\
24  &1.75 &0.02   &2.65 &0.02   &3.92 &0.02  & 9.9 &0.4  &-0.13 &0.01  & 0.02 &0.02  &0.15 &0.02\\
25  &1.87 &0.03   &2.29 &0.04   &3.39 &0.03  & 8.7 &0.6  &-0.32 &0.02  &-0.12 &0.03  &0.20 &0.02\\
26  &1.82 &0.03   &2.57 &0.03   &3.83 &0.03  & 9.0 &0.4  &-0.15 &0.01  & 0.01 &0.03  &0.16 &0.02\\
27  &1.65 &0.03   &2.49 &0.04   &3.97 &0.03  &12.3 &0.7  &-0.26 &0.02  &-0.03 &0.03  &0.23 &0.02\\
28  &1.95 &0.02   &2.61 &0.02   &3.84 &0.02  & 6.9 &0.2  &-0.06 &0.01  & 0.11 &0.02  &0.17 &0.02\\
29  &1.79 &0.01   &2.65 &0.02   &3.84 &0.01  & 9.2 &0.2  &-0.12 &0.01  & 0.02 &0.02  &0.14 &0.02\\
30  &1.72 &0.02   &2.68 &0.02   &4.10 &0.02  &10.3 &0.4  &-0.11 &0.01  & 0.07 &0.02  &0.18 &0.02\\
31  &1.91 &0.02   &2.53 &0.02   &3.78 &0.02  & 7.5 &0.2  &-0.14 &0.01  & 0.04 &0.02  &0.18 &0.02\\
32  &1.79 &0.01   &2.60 &0.01   &4.05 &0.01  & 9.1 &0.2  &-0.13 &0.01  & 0.08 &0.01  &0.21 &0.01\\
33  &1.70 &0.01   &2.69 &0.02   &4.23 &0.01  &10.6 &0.3  &-0.10 &0.01  & 0.10 &0.02  &0.20 &0.02\\
34  &1.78 &0.03   &2.38 &0.03   &3.82 &0.03  &10.0 &0.5  &-0.28 &0.01  &-0.03 &0.02  &0.25 &0.02\\
35  &1.68 &0.02   &2.49 &0.02   &4.05 &0.02  &11.5 &0.4  &-0.24 &0.01  & 0.02 &0.02  &0.26 &0.02\\
36  &1.67 &0.03   &2.60 &0.03   &4.28 &0.03  &11.3 &0.5  &-0.17 &0.02  & 0.10 &0.03  &0.27 &0.02\\
37  &1.88 &0.02   &2.68 &0.02   &4.16 &0.02  & 7.7 &0.2  &-0.03 &0.01  & 0.17 &0.01  &0.20 &0.01\\
38  &1.71 &0.01   &2.71 &0.02   &4.27 &0.01  &10.2 &0.2  &-0.08 &0.01  & 0.13 &0.01  &0.21 &0.01\\
39  &1.60 &0.02   &2.71 &0.02   &4.42 &0.02  &12.3 &0.5  &-0.12 &0.01  & 0.12 &0.02  &0.24 &0.01\\
40  &1.80 &0.02   &2.65 &0.02   &4.14 &0.02  & 8.8 &0.2  &-0.08 &0.01  & 0.13 &0.01  &0.21 &0.01\\
41  &1.69 &0.01   &2.69 &0.01   &4.31 &0.01  &10.6 &0.2  &-0.10 &0.01  & 0.12 &0.02  &0.22 &0.02\\
42  &1.62 &0.02   &2.72 &0.02   &4.46 &0.02  &11.8 &0.3  &-0.10 &0.01  & 0.14 &0.01  &0.24 &0.01\\
43  &1.69 &0.03   &2.45 &0.03   &4.19 &0.03  &11.4 &0.6  &-0.25 &0.02  & 0.06 &0.03  &0.31 &0.02\\
44  &1.64 &0.02   &2.58 &0.02   &4.33 &0.02  &11.9 &0.4  &-0.18 &0.01  & 0.10 &0.01  &0.28 &0.01\\
45  &1.57 &0.02   &2.67 &0.03   &4.54 &0.02  &13.1 &0.6  &-0.14 &0.01  & 0.13 &0.02  &0.27 &0.02\\
46  &1.67 &0.03   &2.73 &0.03   &4.48 &0.03  &10.9 &0.6  &-0.08 &0.02  & 0.17 &0.03  &0.25 &0.02\\
47  &1.65 &0.02   &2.78 &0.02   &4.48 &0.02  &10.9 &0.4  &-0.05 &0.01  & 0.19 &0.02  &0.23 &0.02\\
48  &1.59 &0.03   &2.74 &0.03   &4.69 &0.03  &12.2 &0.6  &-0.08 &0.02  & 0.20 &0.03  &0.28 &0.02\\
49  &1.73 &0.02   &2.69 &0.03   &4.43 &0.02  & 9.8 &0.4  &-0.07 &0.01  & 0.18 &0.03  &0.25 &0.02\\
50  &1.56 &0.01   &2.70 &0.02   &4.56 &0.01  &13.0 &0.4  &-0.12 &0.01  & 0.15 &0.01  &0.27 &0.01\\
51  &1.59 &0.02   &2.81 &0.03   &4.77 &0.02  &12.0 &0.4  &-0.04 &0.01  & 0.23 &0.02  &0.27 &0.02\\
52  &1.61 &0.04   &2.57 &0.04   &4.48 &0.03  &12.7 &0.8  &-0.19 &0.02  & 0.12 &0.03  &0.31 &0.02\\
53  &1.58 &0.03   &2.65 &0.03   &4.62 &0.02  &12.8 &0.6  &-0.14 &0.01  & 0.16 &0.03  &0.30 &0.02\\
54  &1.54 &0.04   &2.69 &0.04   &4.80 &0.04  &13.4 &1.0  &-0.12 &0.02  & 0.20 &0.03  &0.32 &0.03\\
\enddata
\end{deluxetable}

\end{document}